\documentclass[10pt,superscriptaddress,twocolumn,article]{revtex4}

\usepackage[format=plain,justification=justified,singlelinecheck=false,font=small,labelfont=bf,labelsep=space]{caption} 
\usepackage[pdftex, pdfborder={0 0 0}, colorlinks=true, linkcolor=red, urlcolor=blue]{hyperref}
\usepackage{fancyhdr,mathrsfs,times,mathptmx,balance,lastpage}
\usepackage{amsmath,amssymb,amsfonts,bm,graphicx,amsthm}
 \usepackage{url}
\usepackage[toc,page]{appendix}
\usepackage{color,subfigure}
\definecolor{red}{rgb}{0.75,0,0}
\definecolor{blue}{rgb}{0,0,0.75}
\definecolor{turquoise}{rgb}{0.129,0.4516,0.4194}
\definecolor{green}{rgb}{0,0.5,0}
\definecolor{orange}{rgb}{1,.34,.2}
\definecolor{purple}{RGB}{97,0,161}

\newcommand{\Cc}{\mathcal{C}}
\newcommand{\Oc}{\mathcal{O}}

\newcommand{\bbf}{\mathbf{b}}
\newcommand{\ubf}{\mathbf{u}}
\newcommand{\xbf}{\mathbf{x}}
\newcommand{\ybf}{\mathbf{y}}
\newcommand{\Abf}{\mathbf{A}}
\newcommand{\Bbf}{\mathbf{B}}
\newcommand{\Cbf}{\mathbf{C}}
\newcommand{\Lbf}{\mathbf{L}}
\newcommand{\Rbf}{\mathbf{R}}
\newcommand{\Qbf}{\mathbf{Q}}
\newcommand{\Wbf}{\mathbf{W}}
\newcommand{\betabf}{\boldsymbol{\beta}}
\newcommand{\deltabf}{\boldsymbol{\delta}}
\newcommand{\thetabf}{\boldsymbol{\theta}}

\newcommand{\zeros}{\mathbf{0}}

\begin{document}

\title{Models of communication and control for brain networks:\\
distinctions, convergence, and future outlook}

\author{Pragya Srivastava}
\affiliation{Department of Bioengineering, University of Pennsylvania, Pennsylvania, PA 19104, USA}
\author{Erfan Nozari}
\affiliation{Department of Electrical \& Systems Engineering, Pennsylvania, PA 19104, USA}
\author{Jason Z. Kim}
\affiliation{Department of Bioengineering, University of Pennsylvania, Pennsylvania, PA 19104, USA}
\author{Harang Ju}
\affiliation{Department of Neuroscience, Perelman School of Medicine, University of Pennsylvania, Pennsylvania, PA 19104, USA}
\author{Dale Zhou}
\affiliation{Department of Neuroscience, Perelman School of Medicine, University of Pennsylvania, Pennsylvania, PA 19104, USA}
\author{Cassiano Becker}
\affiliation{Department of Bioengineering, University of Pennsylvania, Pennsylvania, PA 19104, USA}
\author{Fabio Pasqualetti}
\affiliation{Department of Mechanical Engineering, University of California, Riverside, 
92521, CA, USA}
\author{Danielle S. Bassett}
\affiliation{Department of Bioengineering, University of Pennsylvania, Pennsylvania, PA 19104, USA}
\affiliation{Department of Physics \& Astronomy, University of Pennsylvania, Pennsylvania, PA 19104, USA}
\affiliation{Department of Electrical \& Systems Engineering, Pennsylvania, PA 19104, USA}
\affiliation{Department of Neurology, Pennsylvania, PA 19104, USA}
\affiliation{Department of Psychiatry, Pennsylvania, PA 19104, USA}
\affiliation{Santa Fe Institute, Santa Fe, NM 87501, USA}

\date{\today}

\begin{abstract}
Recent advances in computational models of signal propagation and routing in the human brain have underscored the critical role of white matter structure. A complementary approach has utilized the framework of network control theory to better understand how white matter constrains the manner in which a region or set of regions can direct or control the activity of other regions. Despite the potential for both of these approaches to enhance our understanding of the role of network structure in brain function, little work has sought to understand the relations between them. Here, we seek to explicitly bridge computational models of communication and principles of network control in a conceptual review of the current literature. By drawing comparisons between communication and control models in terms of the level of abstraction, the dynamical complexity, the dependence on network attributes, and the interplay of multiple spatiotemporal scales, we highlight the convergence of and distinctions between the two frameworks. Based on the understanding of the intertwined nature of communication and control in human brain networks, this work provides an integrative perspective for the field and outlines exciting directions for future work.  
\end{abstract}

\maketitle

\begin{center}
\textbf{\large{Highlights}}
\end{center}

\begin{itemize}
    \item Mathematical models of signal processing and transmission have co-evolved with experimental neuroscience and have been instrumental in enhancing our understanding of emergent `communication dynamics'.
    
    \item The framework of network control theory has increasingly been applied to brain networks to characterize their response to exogeneous or endogeneous stimuli, and to inform the design of intervention strategies to bring desired changes in behaviour. 
    
    \item In this review, we compare the two theoretical approaches in the context of brain networks along the lines of (i) the level of abstraction, (ii) the nature and complexity of models, (iii) the dependence of communication and control measures on network attributes, and (iv) the interplay of different spatiotemporal scales in each.

    \item We discuss outstanding challenges and propose future directions of research that may benefit from combining the two frameworks in the areas of system identification, neuronal decoding, and building biophysically realistic control models.
\end{itemize}

\section{\textbf{Introduction}}
The propagation and transformation of signals among neuronal units that interact via structural connections can lead to emergent communication patterns at multiple spatial and temporal scales. Collectively referred to as `communication dynamics', such patterns reflect and support the computations necessary for cognition \cite{Sporns_2018,Bargmann2013:FromTheConnectome}. Communication dynamics consist of two elements: (i) the dynamics that signals are subjected to, and (ii) the propagation or spread of signals from one neural unit to another. Whereas the former is determined by the biophysical processes that act on the signals, the latter is dictated by the structural connectivity of brain networks. Mathematical models of communication incorporate one or both of these elements to formalize the study of how function arises from structure. Such models have been instrumental in advancing our mechanistic understanding of observed neural dynamics in brain networks \cite{Sporns_2018, BassettNatrev_2018,Misic_2015,Sporns_CurrOp_2013,Bargmann2013:FromTheConnectome, Kopell2014:Beyond,Shen2015:NetworkStructure,Vazquez-Rodriguez2019:Gradients,Cabral2014:Exploring,Bansal2018:Personalized}. 

Building upon the descriptive models of neural dynamics, greater insight can be obtained if one can perturb the system and accurately predict how the system will respond \cite{BassettNatrev_2018}. The step from description to perturbation can be formalized by drawing on both historical and more recent advances in the field of control theory. As a particularly well-developed subfield, the theory of linear systems offers first principles of system analysis and design, both to ensure stability and to inform control \cite{KailathLinearSys}. In recent years, this theory has been applied to the human brain and to non-human neural circuits to ask how inter-regional connectivity can be utilized to navigate the system's state space \cite{Gu_optim_trajec:2017,Tang_2018,Yan2017a,towlson2018celegans}, to explain the mechanisms of endogenous control processes (such as cognitive control) \cite{Gu:2015,cornblath2019sex}, and to design exogenous intervention strategies (such as stimulation) \cite{stiso2019white,khambhati2019functional}. Applicable across spatial and temporal scales of inquiry \cite{Tang_arxiv2019}, the approach has proven useful for probing the functional implications of structural variation in development \cite{Tang_natcomm:2017}, heritability \cite{lee2019heritability}, psychiatric disorders \cite{jeganathan2018fronto,Fisher2014:Electrical}, neurological conditions \cite{bernhardt2019temporal}, neuromodulatory systems \cite{shine2019human} and detection of state transitions \cite{Sabato2012:statedetection,Sabato2011:optimalstatedetection}. Further research in the area of application of network control theory to brain networks can inform neuromodulation strategies \cite{Li2019:Brainstate,Fisher2014:Electrical} and stimulation therapies \cite{Sabato2018:parkinsons}. 

\begin{figure*}
\includegraphics[scale=0.18]{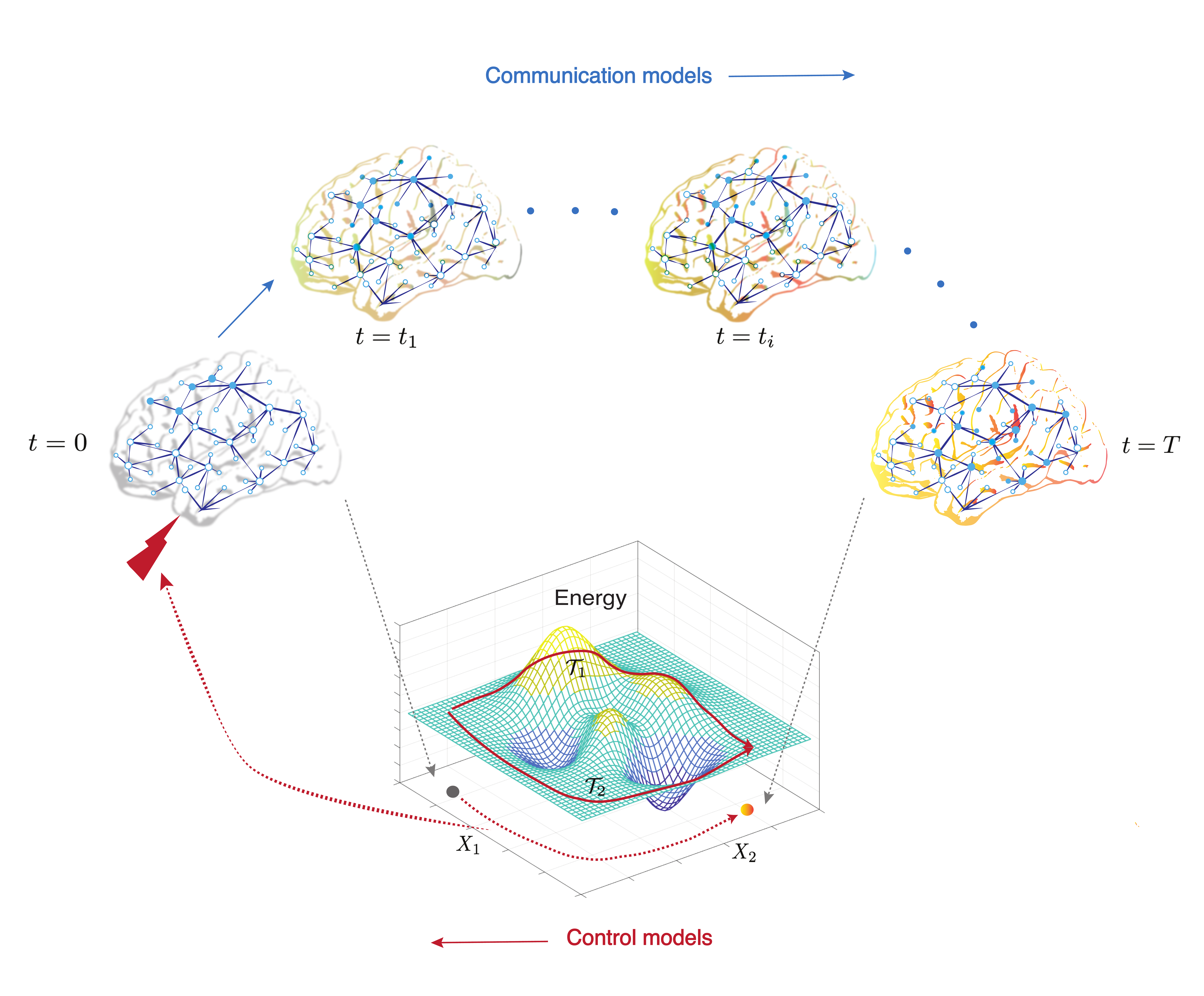}
\caption{
\textbf{The goals of communication and control models share an inverse relationship.} The propagation of an initial stimulus is dictated by the underlying structural connections of the brain network and results in the observed communication dynamics. Stimuli can be external (e.g. transcranial direct current stimulation, sensory stimuli, behavioral therapy, drugs) or internal (e.g. endogenous brain activity, cognitive control strategies). The primary goal of communication models is to capture the evolution of communication dynamics using dynamical models, and to characterize the process of signal propagation using graph-theoretic and statistical measures. In contrast, a fundamental aim in the framework of control theory is to determine the control strategies that would navigate the system from a given initial state to the desired final state. Control signals (shown by the red lightning bolt) move a controllable system along trajectories (shown as a red dotted curve on the state-plane) that connect the initial and final states. Here, the cost of the trajectory is determined by the energetics of the state transition. We show example trajectories $\mathcal{T}_1$ and $\mathcal{T}_2$ on an example energy landscape.\\
}
\label{fig:Fig1}
\end{figure*}

Theoretical frameworks for communication and control share several common features. In communication models, the observed neural activity is strongly influenced by the topology of structural connections between brain regions \cite{Sporns_2018,BassettNatrev_2018}. In control models, the energy injected through exogenous control signals is also constrained to flow along the same structural connections. Thus, the metrics used to characterize communication and control both show strong dependence on the topology of structural brain networks. Interwoven with the topology, the \emph{dynamics} of signal propagation in both the control and communication models involve some level of abstraction of the underlying processes, and dictate the behavior of the system's states. Despite these practical similarities, communication and control models differ appreciably in their goals [Fig. \ref{fig:Fig1}]. Whereas communication models primarily seek to explain the patterns of neural signaling that can arise at rest or in response to stimuli, control theory primarily seeks principles whereby inputs can be designed to elicit desired patterns of neural signaling, under certain assumptions of system dynamics. In other words, at a conceptual level, communication models seek to understand the state transitions that arise from a given set of inputs (including the absence of inputs), whereas control models seek to design the inputs to achieve desirable state transitions.

While relatively simple similarities and dissimilarities are apparent between the two approaches, the optimal integration of communication and control models requires more than a superficial comparison. Here, we provide a careful investigation of relevant distinctions and a description of common ground. We aim to find the points of convergence between the two frameworks, identify outstanding challenges, and outline exciting research problems at their interface. The remainder of this review is structured as follows. First, we briefly review the fundamentals of communication models and network control theory in Sections \ref{sec:section2} and \ref{sec:section3}, respectively. In both sections, we order our discussion of models from simpler to more complex, and we place particular emphasis on each model's spatiotemporal scale. Section \ref{sec:section4} is devoted to a comparison between the two approaches in terms of (i) the level of abstraction, (ii) the complexity of the dynamics and observed behavior, (iii) the dependence on network attributes and, (iv) the interplay of multiple spatiotemporal scales. In Section \ref{sec:section5}, we discuss future areas of research that could combine elements from the two avenues alongside outstanding challenges. Finally, we conclude by summarizing and elucidating the usefulness of combining the two approaches and the implications of such work for understanding brain and behavior.

\section{Communication Models} \label{sec:section2}
\begin{figure*}
\includegraphics[scale=0.18]{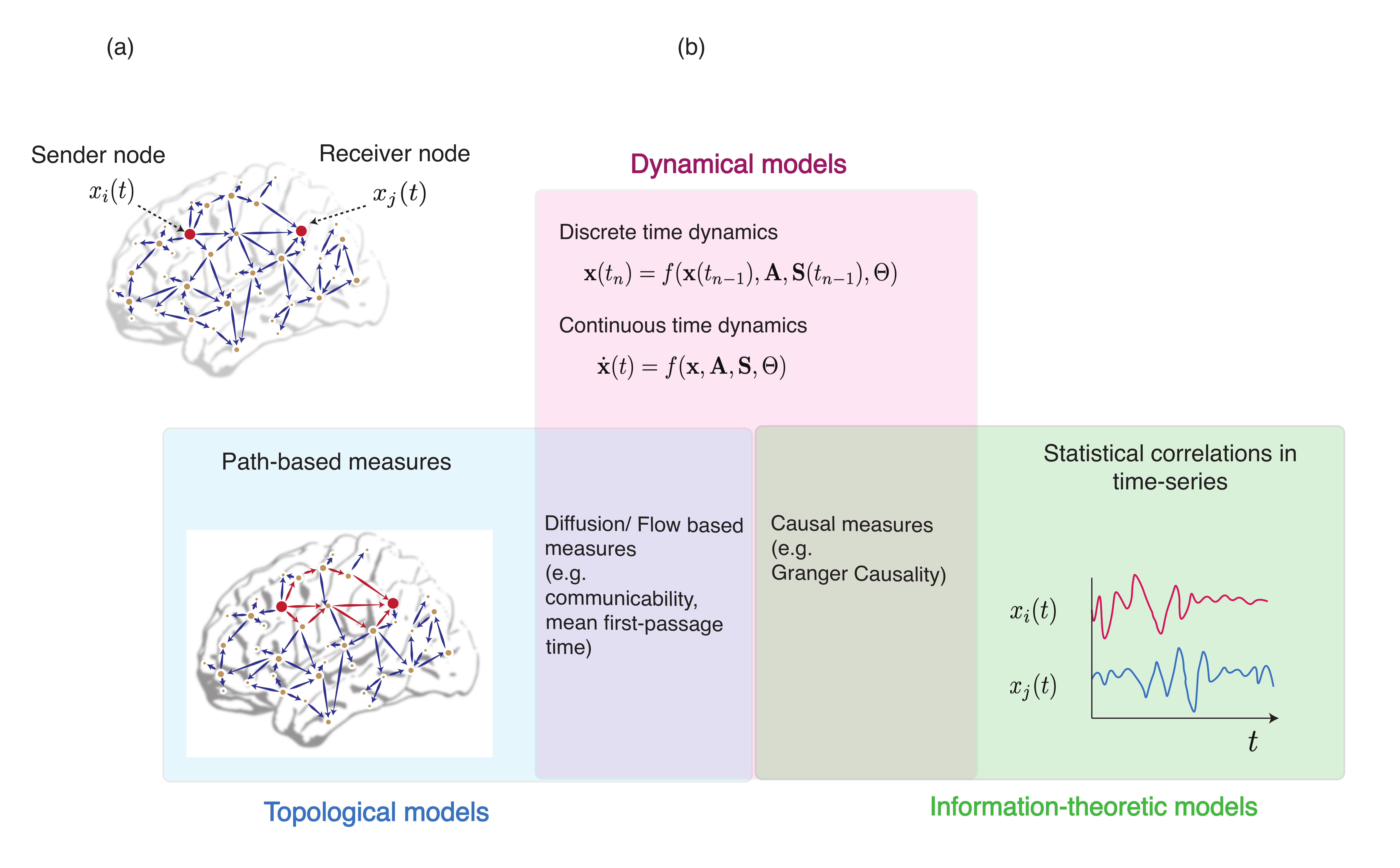}
\caption{\textbf{Models and measures of communication.} (a) A communication event from sender node $i$ to receiver node $j$ causes dependencies in the activity $x_j(t)$ of the $j$-th node on the activity $x_i(t)$ of the $i$-th node. (b) The three classes of mathematical approaches to understanding emergent communication dynamics, as well as potential areas of overlap. Topological models (caerulean) primarily construct measures based on paths or walks (red edges) between communicating nodes. Dynamical models (mauve) can be cast into differential equations (for continuous time dynamics) or difference equations (for discrete time dynamics) that capture dynamic processes governing the propagation of information at a given spatiotemporal scale. Information theoretic models (artichoke) propose measures to compute the degree to which $x_j(t)$ statistically (and sometimes causally) depends upon $x_i(t)$.}
\label{fig:Fig2}
\end{figure*}

In a network representation of the brain, neuronal units are represented as nodes, while inter-unit connections are represented as edges. Such connections can be structural, in which case they are estimated from diffusion imaging \cite{Lazar2010:Mapping}, or can be functional, in which case they are estimated by statistical similarities in activity from functional neuroimaging. When the state of node~$j$ at a given time~$t$ is influenced by the state of node~$i$ at previous time points, a communication channel is said to exist between the two nodes, with node~$i$ being the sender and node~$j$ being the receiver [Fig. \ref{fig:Fig2}(a)]. The set of all communication channels forms the substrate for communication processes. A given communication process can be multiscale in nature: communication between individual units of the network typically leads to the emergence of global patterns of communication thought to play important roles in computation and cognition \cite{Sporns_2018}.

In brain networks, the state of a given node can influence the state of another node precisely because the two are connected by a structural or effective link. This structural constraint on potential causal relations results in patterns of activity reflecting communication among units. Such activity can be measured by techniques such as functional magnetic resonance imaging (fMRI), electroencephalography (EEG), magnetoencephalography (MEG), and electrocorticography (ECoG), among others \cite{Sporns_DCN:2013}. In light of the complexity of observed activity patterns and in response to questions regarding their generative mechanisms, investigators have developed mathematical models of neuronal communication. Such models allow for inferring, relating, and predicting the dependence of measured communication dynamics on the topology of brain networks. 

Communication models can be roughly classified into three types: dynamical, topological, and information theoretic. Dynamical models of communication are generative, and seek to capture the biophysical mechanisms that transform signals and transmit them along structural connections. Topological models of communication propose network attributes, such as measures of path and walk structure, to explain observed activity patterns. Information theoretic models of communication define statistical measures to quantify the interdependence of nodal activity, the direction of communication, and the causal relations between nodes. Several excellent reviews describe these three model types in great detail \cite{Sporns_2018,BassettNatrev_2018, Breakspear_2017, Deco_2008, Hahn_2019}. Thus here we instead provide a rather brief description of the associated approaches and measures, particularly focusing on aspects that will be relevant to our later comparisons with the framework of control theory. 
\\

\subsection{Dynamic models and measures}

Dynamical models of communication aim to capture the biophysical 
mechanisms underlying signal propagation between communicating neuronal units in brain networks. Such models can be defined at various levels of complexity, ranging from relatively simple linear diffusion models to highly non-linear ones. Dynamical models also differ in terms of the spatiotemporal scales of phenomena that they seek to explain. The choice of explanatory scale impacts the precise communication dynamics that the model produces, as well as the scale of collective dynamics that can emerge. 

The general form of a deterministic dynamical model at an arbitrary scale is given by \cite{Breakspear_2017}, 
\begin{equation}
    \frac{d \xbf}{d t} = f(\xbf,\Abf,\ubf,\betabf)\, .
    \label{eq:dyn_eqn}
\end{equation}
\noindent Here, $\xbf$ encodes the state variables that are used to describe the state of the network, $\Abf$ encodes the underlying connectivity matrix, and $\ubf$ encodes the input variables. The functional form of $f$ is set by the requirements (i.e., the expected utility) of the model. For example, at the level of individual neurons communicating via synaptic connections, the conservation law for electric charges (together with model fitting for the gating variables) determines the functional form of $f$ in the Hodgkin-Huxley model \cite{Abbott_1990,Hodgkin_1952}. Similarly, at the scale of neuronal ensembles, other biophysical mechanisms such as the interactions between excitatory and inhibitory populations dictate $f$ in the Wilson-Cowan model \cite{Wilson:1972}. Finally, $\betabf$ encodes other parameters of the model, independent of the connectivity strength $\Abf$. The $\betabf$ parameters can be phenomenological, thereby allowing for an exploration of the whole phase space of possible behaviors; alternatively, the $\betabf$ parameters can be determined from experiments in more data-driven models. In some limiting cases, it may also be possible to derive $\betabf$ parameters in a given model at a particular spatiotemporal scale from complementary models at a finer scale via the procedure of coarse-graining \cite{Breakspear_2017}. 

Fundamentally, dynamical models seek to capture communication of the sort where one unit causes a change in the activity of another unit or shares statistical features with another unit. There is, however, little consensus on precisely how to measure these causal or statistical relations. One of the most common measures is Granger causality~\cite{CWJG:69}, which estimates the statistical relation of unit $x_i$ to unit $x_j$ by the amount of predictive power that the ``past" time series $\{x_i(\tau), \tau < t\}$ of $x_i$ has in predicting $x_j(t)$. While this prediction need not be linear, Granger causality has been historically measured via linear autoregression~\cite{MK-MD-WAT-SLB:01,AK-MM-MK-KJB-SK:03}; see \cite{SLB-AKS:11} for a review in relation to brain networks. 

The use of temporal precedence and lead-lag relationships is also a basis for alternative definitions of causality. In~\cite{GN-AZ-VVN-AS-NK-TB-KM:08}, for instance, the authors propose the phase-slope index, which measures the direction of causal influence between two time series based on the lead-lag relationship between the two signals in the frequency domain. Notably, this relationship can be used to measure the causal effect between neural masses coupled according to the structural connectome~\cite{CJS-ECWVS:12}. Because not all states of a complex system can often be measured, several studies have opted to first reconstruct (equivalent) state trajectories via time delay embedding~\cite{FT:81,CRS:06} before measuring predictive causal effects~\cite{GS-RM-HY-CH-ED-MF-SM:12,DH-EL-MS-KRP:17}. Finally, given the capacity to perturb the states or even parameters of the network (either experimentally or in simulations), one can observe the subsequent changes in other network states that occur, and thereby discover and measure causal effects \cite{DAS:14,DAS:18}.

\subsection{Topological models and measures}\label{subsec:topological}

The potential for communication between two brain regions, each represented as a network node, is dictated by the paths that connect them. It has been thought that long routes demand high metabolic costs and sustain marked delays in signal propagation \cite{bullmore2012economy}. Thus, the presence and nature of \emph{shortest} paths through a network are commonly used to infer the efficiency of communication between two regions \cite{Sporns_2018}. If the shortest path length between node $i$ and node $j$ is denoted by $d(i,j)$ \cite{Latora_2001} then the \emph{global efficiency} through a network is defined as the mean of the inverse shortest path lengths $\epsilon_{ij} = \frac{1}{d(i,j)}$ \cite{Latora_2001,Ek_2015}. Although measures based on shortest paths have been widely used, their relevance to the true system has been called into question for three reasons. First, systems that route information exclusively through shortest paths are vulnerable to targeted attack of the associated edges \cite{Sporns_2018}; yet, one might have expected brains to have evolved to circumvent this vulnerability, for example by also using non-shortest paths for routing. Second, a sole reliance on shortest-path routing implies that brain networks have non-optimally invested a large cost in building alternative routes which essentially are not used for communication. Third, the ability to route a signal by the shortest path appears to require the signal or brain regions to have biologically implausible knowledge of the global network structure. These reasons have motivated the development of alternative measures, such as the number of parallel paths or edge-disjoint paths between two regions \cite{Sporns_2018}; systems using such diverse routing strategies can attain greater resilience of communication processes~\cite{avena2019spectrum}. The resilience of inter-regional communication in brain networks is a particularly desired feature since fragile networks have been found 
to be associated with neurological disorders such as epilepsy \cite{Adam2017:fragilityepileptic,sritharan2014fragility,ehrens2015closed}. 

The assumption of information flow through all paths available between two regions leads to the notion of \emph{communicability}. By denoting the adjacency matrix $\Abf$, we can define the communicability between node $i$ and node $j$ as the weighted sum of all walks starting at node $i$ and ending at node $j$ \cite{Estrada2008,Estrada_2012}: 

\begin{equation}
    \label{eq:communicability}
    G_{ji} = \overset{\infty}{\underset{k =0}{ \Sigma}} c_k (\Abf^k)_{ji}\, ,
\end{equation}

\noindent where $\Abf^k$ denotes the $k-$th power of $\Abf$, and $c_k$ are appropriately selected coefficients that both ensure that the series is convergent and assign smaller weights to longer paths. If the entries of $\Abf$ are all nonnegative (which is the context in which communicability is mainly used), then $G_{ji}$ is also real and non-negative. Out of several choices that can be made, a particularly insightful one is $c_k = \frac{1}{k!}$. The resulting communicability, also known as the exponential communicability $G_{ji} = (e^\Abf)_{ji}$, allows for interesting analogies to be drawn with the thermal Green's function and correlations in physical systems \cite{Estrada2008, Estrada_2012}. Additionally, since $(\Abf^k)_{ji}$ directly encodes the weighted paths of length $k$ from node $i$ to node $j$, one can conveniently study the path length dependence of communication. Exponential communicability is also similar to the impulse response of the system, a familiar notion in control theory which we further explore in Section \ref{sec:section4}.  

Another flow-based measure of communication efficiency is the \emph{mean first-passage time}, which quantifies the distance between two nodes when information is propagated by diffusion. Similar to the \emph{global efficiency}, the \emph{diffusion efficiency} is the average of the inverse of the mean first-passage time between all pairs of network nodes. Interestingly, systems that evolve under competing constraints for diffusion efficiency and routing efficiency can display a diverse range of network topologies \cite{Sporns_2018}. Note that these global measures of communication efficiency only provide an upper bound on the assumed communicative capacity of the network; in networks with significant community or modular structures, other architectural attributes such as the existence and inter-connectivity of highly connected hubs are determinants of the integrative capacity of a network which global measures of communication efficiency fail to capture accurately \cite{Sporns_CurrOp_2013}.

Network attributes that determine an efficient propagation of externally induced or intrinsic signals may inform generative models of brain networks both in health and disease \cite{vertes2012simple}. Moreover, such attributes can inform the choice of control inputs targeted to guide brain state transitions; we discuss this convergence in Section \ref{sec:section4}. Further, quantifying communication channel capacity calls for the use of information theory, which we turn to now.

\subsection{Information theoretic models and measures}

Information theory and statistical mechanics have been used to define several measures of information transfer such as transfer entropy and Granger causality. Such measures are built upon the fact that the process of signal propagation through brain networks results in collective time-dependent activity patterns of brain regions which can be measured as time-series. Entropic measures of communication aim to find statistical dependencies between such time-series to infer the amount and direction of information transfer. The processes underlying the observed time-series are typically assumed to be Markovian, and measures of statistical dependence are calculated in a manner that reflects causal dependence. For this reason, the causal measures of communication proposed in the information theoretic approach share similarities with those used in dynamical causal inference \cite{Valdes-Sosa_2011}. 

A central quantity in information theory is the Shannon entropy, which measures the uncertainty in a discrete random variable $I$ that follows the distribution $p(i)$ and is given by  $H(I) = - \underset{i}{\Sigma} p(i) \log(p(i))$. One measure of statistical inter-dependency between two random variables $I$ and $J$ is their mutual information, $ M_{IJ} = \Sigma p(i,j) \frac{\log(p(i,j))}{\log{p(i)}, \log{p(j)}}$, where $p(i,j)$ is their joint distribution and $p(i)$ and $p(j)$ are its marginals. Since mutual information is symmetric, it fails to capture the direction of information flow between two \emph{processes} (sequences of random variables) \cite{Schreiber_2000}. 

To address this limitation, the measure of transfer entropy was proposed to capture the directionality of information exchange \cite{Schreiber_2000}. Transfer entropy takes into account the transition probability between different \emph{states}, which can be the result of a stochastic dynamic process (similar to Eq.~\eqref{eq:dyn_eqn} but with a stochastic $\ubf$) and obtained from the time series of activities of brain regions through imaging techniques. To measure the direction of information transfer between processes $I$ and $J$, the notion of mutual information is generalized to the mutual information rate. The transfer entropy between processes $I$ and $J$ is given by \cite{Schreiber_2000}: 

\begin{equation}
    T_{J \rightarrow I} = \Sigma p (i_{n+1}, i_n^{(k)},j_n^{(l)}) 
    \log \frac{p (i_{n+1} \vert i_n^{(k)},j_n^{(l)})}{ p (i_{n+1} \vert i_n^{(k)})}\, ,
\end{equation}

\noindent where processes $I$ and $J$ are assumed to be stationary Markov processes of order $k$ and $l$, respectively. The quantity $i_n^{(k)}$ ($j_n^{(l)}$) denotes the state of process $I$($J$) at time $n$ while $p((i_{n+1} \vert i_n^{(k)})$ denotes the transition probability to state $i_{n+1}$ at time $n+1$, given knowledge of the previous $k$ states. The quantity $ p (i_{n+1}\vert i_n^{(k)},j_n^{(l)} )$ is the same as $p(i_{n+1} \vert i_n^{(k)})$ if the process $J$ does not influence the process $I$. 

Similar to Granger causality, transfer entropy has been extensively used to compute the statistical interdependence of dynamic processes and to infer the directionality of information exchange. Later studies have sought to combine these two measures into a single framework by defining the \emph{multi-information}. This approach takes into account the statistical structure of the whole system and of each subsystem, as well as the structure of the interdependence between them \cite{Chicharro_2012}. Such methods complement the topological and dynamical models to provide a unique perspective on communication, by quantifying information content and transformation.\\

\subsection{Communication models across spatiotemporal scales} 

Whether considering models that are dynamical, topological, or information theoretic, we must choose the identity of the neural unit that is performing the communication. Individual neurons form basic units of computation in the brain, which communicate with other neurons via synapses. One particularly common model of communication at this cellular scale is the Hodgkin-Huxley model, which identifies the membrane potential as the state variable whose evolution is determined by the conservation law for electric charge \cite{Hodgkin_1952}. Simplifications and dimensional reductions of the Hodgkin-Huxley model have led to related models such as the Fitzhugh-Nagumo model, which is particularly useful for studying the resulting phase space \cite{Fitzhugh_1961,Abbott_1990}. Further simplifications of the neuronal states to binary variables have facilitated detailed accounts of network-based interactions such as those provided by the Hopfield model \cite{Abbott_1990,BassettNatrev_2018}. Collectively, despite all capturing the state of an individual neuron, these models differ from one another in the biophysical realism of the chosen state variables: the on/off states in the Hopfield model are arguably less realistic than the membrane potential state in the Hodgkin-Huxley model. 

When considering a large population of neurons, a set of simplified dynamics can be derived from those of a single neuron using the formalism and tools from statistical mechanics \cite{Deco_2008,Breakspear_2017,Abbott_1990}. The approximations prescribed by the laws of statistical mechanics -- such as, for example, the diffusion approximation in the limit of uncorrelated spikes in neuronal ensembles -- have led to the Fokker-Planck equations for the probability distribution of neuronal activities. From the evolution of such probability distributions, one can derive the dynamics of the moments, such as the mean firing rate and variance \cite{Deco_2008,Breakspear_2017}. Several models of neuronal ensembles exist that exhibit rich collective behavior such as synchrony \cite{Palmigiano_2017,Vuksanovic2015:Dynamic}, oscillations \cite{Fries_2005,Kopell2010:GammaAndTheta}, waves \cite{Muller_2018,Roberts_2019}, and avalanches \cite{Beggs_2003}, each supporting different modes of communication. In the limit where the variance of neuronal activity over the ensemble can be assumed to be constant (e.g., in the case of strong coherence), the Fokker-Planck equation leads to neural mass models \cite{Breakspear_2017}. Relatedly, the Wilson-Cowan model is a mean-field model for interacting excitatory and inhibitory populations of neurons \cite{Wilson:1972}, and has significantly influenced the subsequent development of theoretical models for brain regions \cite{Destexhe:2009,Kameneva2017:NeuralMass}. At scales larger than that of neuronal ensembles, brain dynamics can be modelled by coupling neural masses, Wilson-Cowan oscillators, or Kuramato oscillators according to the topology of structural connectivity \cite{Breakspear_2017,Muller_2018,Roberts_2019,Palmigiano_2017,Sanz-Leon2015:Mathematical}. Collectively, these models provide a powerful way to theoretically and computationally generate the large scale temporal patterns of brain activity which can be explained by the theory of dynamical systems. 

When changing models to different spatiotemporal scales, we must also change how we think about communication. While communication might involve induced spiking at the neuronal scale, it may also involve phase lags at the population scale. Dynamical systems theory provides a powerful and flexible framework to determine the emergent behavior in dynamic models of communication. As we saw in Eq.~\eqref{eq:dyn_eqn}, the evolution of the system is represented by a trajectory in the phase space constructed from the system's state variables. A critical notion from this theory has been that of attractors, namely, stable patterns in this phase space to which phase trajectories converge. The range of emergent behavior exhibited by the dynamical system such as steady states, oscillations, and chaos, is thus determined by the nature of its attractors which can be stable fixed points, limit cycles, quasi-periodic, or chaotic.  Oscillations, synchronization, and spiral or travelling wave solutions that result from dynamical models match with the patterns observed in brain networks, and have been proposed as the mechanisms contributing to cross-regional communication in brain \cite{Roberts_2019,Buelhmann_2010,Deco_2011,Besserve_2015,Rubino_2006}.

The class of communication models that generate oscillatory solutions holds an important place in models of brain dynamics~\cite{breakspear2010generative}. Numerous classes of nonlinear models at both the micro- and macro-scale exhibit oscillatory solutions, and they can be broadly classified into periodic (limit cycle), quasi-periodic (tori), and chaotic~\cite{Breakspear_2017}. Synchronization in the activity of spiking neurons is an emergent feature of neural systems that appears to be particularly important for a variety of cognitive functions \cite{Bennett:2004}. This fact has motivated efforts to model brain regions as interacting oscillatory units, whose dynamics are described by, for example, the Kuramoto model for phase oscillators. In its original form, the equation for the phase variable~$\theta_i(t)$ of the $i-$th Kuramoto oscillator is given by \cite{YK:03,Acebron:2005}

 \begin{align}
 \label{eq:kuramoto}
     \dot \theta_i(t) = \omega_i + \sum_{j = 1}^n A_{ij} \sin(\theta_j(t) - \theta_i(t)) ,
    \end{align}

\noindent where $\omega_i$ denotes the natural frequency of oscillator $i$, which depends upon its local dynamics and parameters, and $A_{ij}$ denotes the net connection strength of oscillator~$j$ to oscillator~$i$. Phase oscillators generally and the Kuramoto model specifically have been widely used to model neuronal dynamics~\cite{breakspear2010generative}. The representation of each oscillator by its phase (which critically depends upon the weak coupling assumption~\cite{GBE-NK:90}) makes it particularly tractable to study synchronization phenomena~\cite{SB-VL-YM-MC-DUH:06,NC-MWS:09,TM-GB-DSB-FP:19}. Generalized variants of the Kuramoto model have also been proposed and studied in the context of neuronal networks \cite{Cumin:2007}.\\

\section{Control Models} \label{sec:section3}

\begin{figure*}
\includegraphics[scale=0.24]{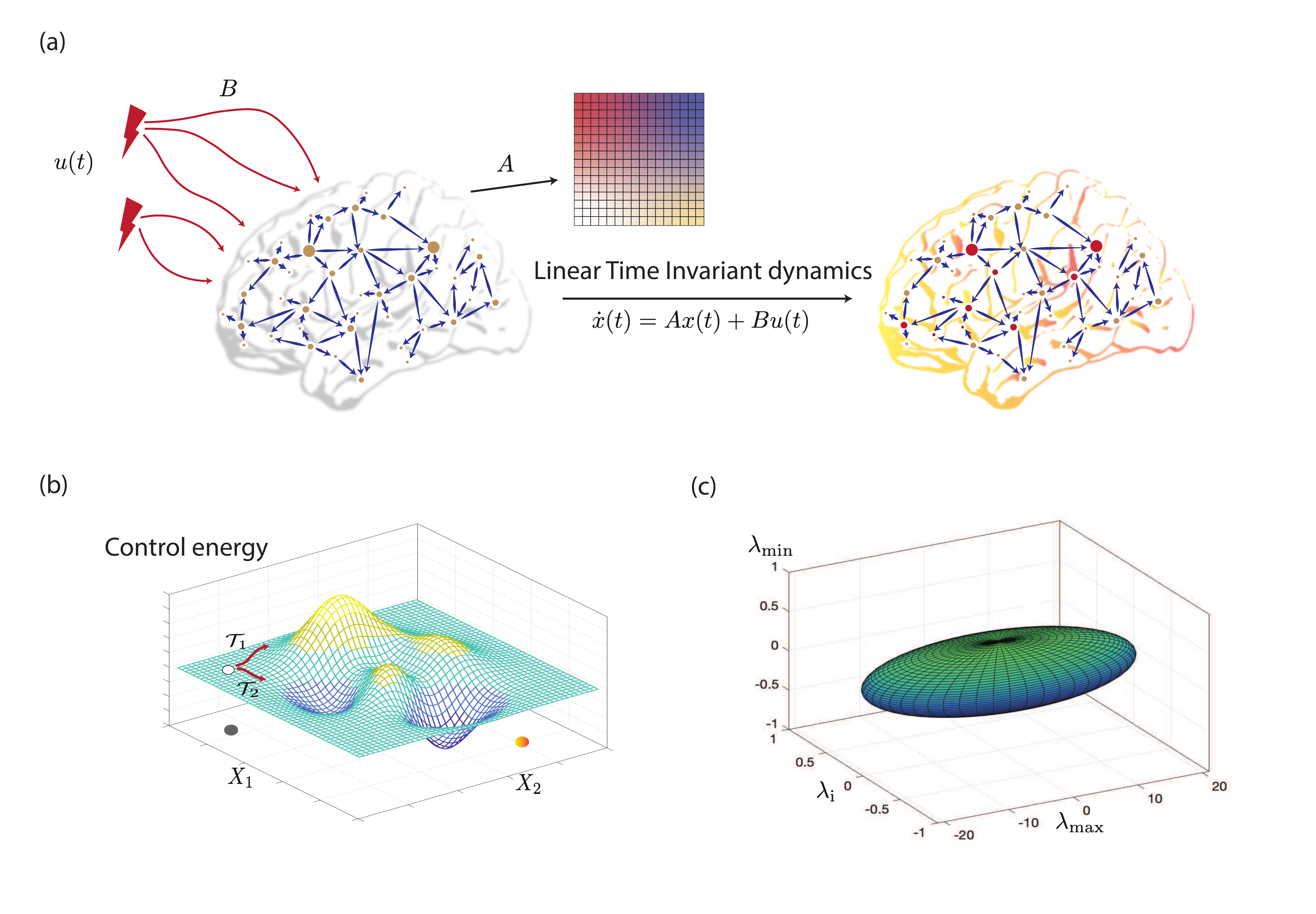}
\caption{\textbf{Control theory applied to brain networks.} Control theory seeks to determine and quantify the controllability and observability properties of a given system. A system is controllable when a control input $u(t)$ is guaranteed to exist to navigate the system from a given initial state to a desired final state in a specified span of time. (\textbf{a}) We begin by encoding a brain network in the adjacency matrix denoted by $A$. Then, control signals $u(t)$ act on the network via the input matrix $B$, leading to the evolution of the system's state to a desired final state according to some dynamics. The most common dynamics studied in this context is a linear time invariant dynamics. Whether the system can be navigated between two arbitrary states in a given time period is determined by a full-rank condition on the controllability matrix.  (\textbf{b}) The control energy landscape dictates the availability and difficulty of transitions between distinct system states. For a controllable system, several trajectories can exist which connect the initial and final states. An optimum trajectory is then determined using the notion of optimal control. (\textbf{c}) The eigenvalues of the inverse Gramian matrix quantify the ease of moving the system along eigen-directions that span the state-space and form an $N-$ ellipsoid whose surface reflects the control energy to make unit changes in the state of the system along the corresponding eigen-direction. Here we show an ellipsoid constructed from the maximum, the minimum, and an intermediate eigenvalue of the Gramian for an example regular graph with $N=400$ nodes and degree $l=40$. The initial state has been taken to be at the origin, and the final state is a random vector of length $N$ with unit norm. Commonly used metrics of controllability such as the average controllability, can be constructed from the eigenvalues of the Gramian.}
\label{fig:Fig3}
\end{figure*}

While the study of communication in the neural systems has developed hand-in-hand with our understanding of the brain, the study of control dynamics in (and upon) the brain is rather young and still in early stages of development. In this section we review some of the basic elements of control theory that will allow us in later sections to elucidate the relationships between communication and control in brain networks. 

\subsection{The theory of linear systems}
The simplicity and tractability of linear time-invariant (LTI) models have sparked significant interest in the application of linear control theory to neuroscience \cite{KailathLinearSys,Tang_2018}. LTI systems are most commonly studied in state space, and their simplest form is finite dimensional, deterministic, without delays, and without instantaneous effects of the input on the output. Such a continuous-time LTI system is described by the algebraic-differential equation
\begin{subequations}\label{eq:LTI}
\begin{align}
    \label{eq:LTI-state} \frac{d}{dt} \xbf(t) &= \Abf \xbf(t) + \Bbf \ubf(t)
    \\
    \label{eq:LTI-output} \ybf(t) &= \Cbf \xbf(t).
\end{align}
\end{subequations}
\noindent Here, Eq.~\eqref{eq:LTI-state} is a special case of Eq.~\eqref{eq:dyn_eqn} (with the input matrix $\Bbf$ corresponding to $\betabf$), while the output vector $\ybf$ now allows for a distinction between the internal, latent state variables $\xbf$ and the external signals that can be measured, say, via neuroimaging. Each element $C_{ij}$ of the matrix $\Cbf$ thus describes the loading of the $i$-th measured signal on the activity level of the $j$-th brain region. Note that the number of states, inputs, and outputs need not be the same, in which case $\Bbf$ and $\Cbf$ are not square matrices. 

At the macroscale where linear models are most widely used, the state vector $\xbf$ often contains as many elements as the number of brain (sub)regions of interest with each element $x_i(t)$ representing the \emph{activity level} of the corresponding region at time $t$, for example corresponding to the mean firing rate or local field potential. The elements of the vector $\ubf$ are often more abstract and can model either internal or external sources. An example of an internal source would be a cognitive control signal from frontal cortex, whereas an example of an external source would be neurostimulation \cite{Gu:2015,cornblath2019sex,cui2020optimization,sritharan2014fragility,ehrens2015closed}. While a formal link between these internal or external sources and the model vector $\ubf$ is currently lacking, it is standard to let $\int_0^T \vert \ubf (t)\vert^2 dt$ represent the net \emph{energy}. The matrix $\Bbf$ is often binary, with one nonzero entry per column, and encodes the spatial distribution of the input channels to brain regions. 

Owing to the tractability of LTI systems, the state response of an LTI system (i.e. $\xbf(t)$) to a given stimulus $\ubf(t)$ can be analytically obtained as:

\begin{align}\label{eq:LTI-sol}
    \xbf(t) = e^{\Abf t} \xbf(0) + \int_0^t e^{\Abf(t - \tau)} \Bbf \ubf(\tau) d \tau .
\end{align}
In this expression, the matrix exponential $e^{\Abf t}$ has a special significance. If $\xbf(0) = \zeros$, and if $\ubf_i(t)$ is an impulse (i.e., a Dirac delta function) for some $i$, and if the remaining input channels are kept at zero, then Eq.~\eqref{eq:LTI-sol} simplifies to the system's \emph{impulse response}
\begin{align}
    \label{eq:impulse}
    \xbf(t) = e^{\Abf t} \bbf_i\, ,
\end{align}
where $\bbf_i$ is the $i$-th column of $\Bbf$. Clearly, the impulse response has close ties to the \emph{communicability} property of the network introduced in Section~\ref{subsec:topological}. We discuss this relation further in Section \ref{sec:section4}, where we directly compare communication and control. 

\subsection{Controllability and observability in principle}\label{subsec:con-obs}

One of the most successful applications of linear control theory to neuroscience lies in the evaluation of controllability. If the input-state dynamics (Eq.~\eqref{eq:LTI-state}) is controllable, it is possible to design a control signal $\ubf(t), t \ge 0$ such that $\xbf(0) = \xbf_0$ and $\xbf(T) = \xbf_f$ for any \emph{initial state} $\xbf_0$, \emph{final state} $\xbf_f$, and \emph{control horizon} $T > 0$. In other words, a (continuous-time LTI) system is controllable if it can be controlled from any initial state to any final state in a given amount of time; notice that controllability is independent of the system's output. Using standard control-theoretic tools, it can be shown that the system Eq.~\eqref{eq:LTI-state} is controllable if and only if the \emph{controllability matrix} $\Cc = \begin{bmatrix} \Bbf & \Abf\Bbf & \cdots & \Abf^{n-1}\Bbf \end{bmatrix}$ has full rank $n$, where $n$ denotes the dimension of the state \cite{KailathLinearSys}.

The notion of full-state controllability discussed above can at times be a strong requirement, particularly as the size of the network (and therefore the dimension of the state space) grows. If it happens that a system is \emph{not} full-state controllable, the control input $\ubf(t)$ can still be designed to steer the state in \emph{certain directions}, despite the fact that not every state transition is achievable. In fact, we can precisely determine the directions in which the state can and cannot be steered using the input $\ubf(t)$. The former, called the \emph{controllable subspace}, is given by the range space of the controllability matrix $\Cc$: all directions that can be written as a linear combination of the columns of $\Cc$. It can be shown that the state can be arbitrarily steered within the controllable subspace, similar to a full-state controllable system \cite[\S6.4]{CTC:98}. Recall, however, that the rank of $\Cc$ is necessarily less than $n$ for an uncontrollable system, and so is the dimension of the controllable subspace. If this rank is $r < n$, we then have an $n - r$ dimensional subspace, called the \emph{uncontrollable subspace}, which is orthogonal to the controllable one. In contrast to our full control over the controllable subspace, the evolution of the system is completely autonomous and independent of $\ubf(t)$ in the uncontrollable subspace \cite{KailathLinearSys}.

Dual to the notion of controllability is that of \emph{observability}, which has been explored to a lesser degree in the context of brain networks. Whereas an output can be directly computed when the input and initial state are specified (Eq.~\eqref{eq:LTI}), the converse is not necessarily true; it is not always possible to solve for the state from input-output measurements. The property that characterizes and quantifies the possibility of determining the state from input-output measurements is termed \emph{observability}, and can be understood as the possibility to invert the state-to-output map (Eq.~\eqref{eq:LTI-output}), albeit over time. Interestingly, the input signal $\ubf(t)$ and matrix $\Bbf$ are irrelevant for observability. Moreover, the system Eq.~\eqref{eq:LTI} is observable if and only if its \emph{dual system} $d\xbf(t)/dt = \Abf^T \xbf(t) + \Cbf^T \ubf(t)$ is \emph{controllable} (here, the superscript $^T$ denotes the transpose). This duality allows us to, for instance, easily determine the observability of Eq.~\eqref{eq:LTI} by checking whether the \emph{observability matrix} $\Oc = \begin{bmatrix} \Cbf^T & \Abf^T \Cbf^T & \cdots & (\Abf^T)^{n-1} \Cbf^T \end{bmatrix}^T$ has full rank. The notion of observability may be particularly relevant to the measurement of neural systems, and we discuss this topic further in Sections \ref{sec:section4} and \ref{sec:section5}. 

\subsection{Controllability in practice}\label{subsec:con-prac}
Once a system is determined to be controllable in principle, the next natural question is how to design a control signal $\ubf(t)$ that can move the system between two states. Although the existence of at least one such signal is guaranteed by controllability, this control signal and the resulting system trajectory may not be unique; for instance, an arbitrary intermediate point can be reached in $T/2$ time and then the final state can be reached in the remaining time (both due to controllability). This non-uniqueness of control strategies leads to the problem of \emph{optimal control}; that is, designing \emph{the best} control signals that achieve a desired state transition, according to some criterion of optimality. The simplest and most commonly used criterion is the control energy defined as

\begin{align}\label{eq:min-energy}
    \int_0^T \|u(t)\|^2 d t = \int_0^T \sum_{j = 1}^m u_j(t)^2 dt\, ,
\end{align}

\noindent where $\|\cdot\|$ denotes the Euclidean norm. The corresponding control signal that minimizes~\eqref{eq:min-energy} is thus referred to as the \emph{minimum energy control}. Owing to the tractability of LTI systems, this control signal and its total energy can be found analytically~\cite{DEK:04}.

While certainly useful, the minimum energy criterion (Eq.~\eqref{eq:min-energy}) has a number of limitations. In particular, the energy of all the control channels are weighted equally. Further, the state is allowed to become arbitrarily large between the initial and final times. These limitations have motivated the more general linear-quadratic regulator (LQR) criterion
\begin{equation}
\begin{aligned}\label{eq:lqr}
        \int_0^T \left(\sum_{j = 1}^m R_j u_j(t)^2 + \sum_{i = 1}^n Q_i x_i(t)^2\right) dt\\ = \int_0^T \left[\ubf(t)^T \Rbf \ubf(t) + \xbf(t)^T \Qbf \xbf(t)\right] dt\, ,
\end{aligned}
\end{equation}

\noindent where $R_j$ and $Q_i$ are positive weights forming the diagonal entries of the matrices $\Rbf$ and $\Qbf$, respectively, and $^T$ denotes the transpose operator. Whereas the first term in Eq.~\eqref{eq:lqr} expresses the cost of control as in Eq.~\eqref{eq:min-energy}, the second term introduces a cost on the trajectory in state-space. This general form poses a trade-off between the two costs, and is particularly relevant in cases where some regions of state space are more preferred than others. By selecting the entries of $\Qbf$ to be large relative to $\Rbf$, for instance, the resulting control will ensure that the state remains close to $\zeros$. The second term in Eq.~\eqref{eq:lqr} can further be generalized to introduce a preferred trajectory in the state-space by replacing $\xbf(t)$ by $\xbf(t) - \xbf^\ast(t)$ where $\xbf^\ast(t)$ denotes the preferred trajectory. An analytical solution can also be found for the control signals minimizing the above generalized energy. Notably, the cost function Eq.~\eqref{eq:lqr} has recently proven fruitful in the study of brain network architecture and development \cite{Gu_optim_trajec:2017,Tang_natcomm:2017}.

Another central quantity of interest in characterizing the controllability properties of an LTI system is the Gramian matrix which, for continuous-time dynamics, is given as

\begin{equation}
\label{eq:Cont-LTI-Gram}
\Wbf_T = \int_0^T e^{\Abf t} \Bbf \Bbf^T e^{\Abf^T t} dt .
\end{equation}

\noindent The invertibility of the Gramian matrix, equivalently to the full-rank condition of the controllability matrix, ensures that the system is controllable. Further, the eigen-directions (eigenvectors) of the Gramian corresponding to its nonzero (positive) eigenvalues form a basis of the state subspace that is reachable by the system (Fig.~\ref{fig:Fig3}(c)) \cite{Liu2011a,Lewis_OptimControl}, even when the Gramian is not invertible (note the relation with the controllable and uncontrollable subspaces discussed above). Intuitively then, the eigenvalues of the Gramian matrix quantify the ease of moving the system along corresponding eigen-directions. Various efforts have thus been made to condense the $n$ eigenvalues of the Gramian into a single, scalar \emph{controllability metric}, such as the average controllability and control energy (see below) \cite{KailathLinearSys,Tang_2018,Gu_optim_trajec:2017,FabioMetrics_2014}. 

Using the controllability Gramian, it can in fact be shown that the energy~\eqref{eq:min-energy} of the mininum-energy control is given by (assuming $\xbf(0) = \zeros$ for simplicity)
\begin{equation}
\label{eq:E_as_gramian}
    E = \xbf_f^T \Wbf_T^{-1} \xbf_f\, ,
\end{equation}
\noindent where $\xbf_f$ denotes the final state. The framework of minimum energy control and controllability metrics have recently been applied to brain networks, see, e.g. \cite{Tang_natcomm:2017,Tang_2018,Gu:2015,Gu_optim_trajec:2017}. 
This framework further opens up interesting questions about its implications for control and the response of brain networks to stimuli; specifically, one might with to determine the physical interpretation of controllability metrics in brain networks and how they can inform optimal intervention strategies. We revisit this point while discussing the utility of communication models in addressing some of these questions in Section~\ref{sec:section4}-B.

\subsection{Generalizations to time-varying and non-linear systems}

Used most often due to its simplicity and analytical tractability, the LTI model of system's dynamics limits the temporal behavior that can be exhibited by the system to the following three types: exponential growth, exponential decay, and sinusoidal oscillations. In contrast, the brain exhibits a rich set of dynamics encompassing many other types of behaviors. Numerical simulation studies have sought to understand how such rich dynamics, occurring atop a complex network, respond to perturbative signals such as stimulation \cite{muldoon2016stimulation,papadopoulos2020relations}.  Yet, to more formally bring control theoretic models closer to such dynamics and associated responses, the framework must be generalized to include non-linearity and/or time-dependence. The first step in such a generalization is the Linear Time-Varying (LTV) system
\begin{subequations}\label{eq:ltv}
\begin{align}
    \frac{d}{dt} \xbf(t) &= \Abf(t) \xbf(t) + \Bbf(t) \ubf(t)
    \\
    \ybf(t) &= \Cbf(t) \xbf(t) .
\end{align}
\end{subequations}
\noindent Notably, a generalization of the optimal control problem (Eq.~\eqref{eq:lqr}) to LTV systems is fairly straightforward \cite{DEK:04}. But, unlike LTI systems (Eq.~\eqref{eq:LTI-sol}), it is generically not possible to solve for the state trajectory of an LTV system analytically. However, if the state trajectory can be found for $n$ linearly independent initial states, then it can be found for any other initial state due to the property of linearity. In this case, moreover, many of the properties of LTI systems can be extended to LTV systems~\cite{CTC:98}, including the simple rank conditions of controllability and observability \cite{LMS-HEM:67}.

Moving beyond the time-dependence addressed in LTV systems, one can also consider the many nonlinearities present in real-world systems. In fact, the second common generalization of LTI systems (Eq.~\eqref{eq:LTI}) is to nonlinear control systems which, in continuous time, have the general state-space representation:
\begin{subequations}\label{eq:nonlin}
\begin{align}
    \frac{d}{dt} \xbf(t) &= f(\xbf(t), \ubf(t), t)
    \\
    \ybf(t) &= h(\xbf(t), t) .
\end{align}
\end{subequations}
\noindent The time-dependence in $f$ and $h$ may be either explicit or implicit via the time-dependence of $\xbf$ and $\ubf$, resulting in a \emph{time-varying} or \emph{time-invariant} nonlinear system, respectively. 

Before proceeding to truly nonlinear aspects of Eq.~\eqref{eq:nonlin}, it is instructive to consider the relationship between these dynamics and the linear models described above (Eqs.~\eqref{eq:LTI} and~\eqref{eq:ltv}). Assume that for a given input signal $\ubf_0(t)$, the solution to Eq.~\eqref{eq:nonlin} is given by $\xbf_0(t)$ and $\ybf_0(t)$. As long as the input $\ubf(t)$ to the system remains close to $\ubf_0(t)$ for all time, then $\xbf(t)$ and $\ybf(t)$ also remain close to $\xbf_0(t)$ and $\ybf(t)$, respectively. Therefore, one can study the dynamics of small perturbations $\deltabf_x(t) = \xbf(t) - \xbf_0(t)$, $\deltabf_u(t) = \ubf(t) - \ubf_0(t)$, and $\deltabf_y(t) = \ybf(t) - \ybf_0(t)$ instead of the original state, input, and output. Using a first order Taylor expansion, it can immediately be seen that these signals approximately satisfy
\begin{subequations}\label{eq:ltv-linearized}
\begin{align}
    \frac{d}{dt} \deltabf_x(t) &= \Abf(t) \deltabf_x(t) + \Bbf(t) \deltabf_u(t)
    \\
    \deltabf_y(t) &= \Cbf(t) \deltabf_x(t) ,
\end{align}
\end{subequations}
\noindent which is an LTV system of the form given in Eq.~\eqref{eq:ltv}. In these equations, $\Abf(t) = \frac{\partial f(\xbf_0(t), \ubf_0(t), t)}{\partial \xbf_0(t)}$, $\Bbf(t) = \frac{\partial f(\xbf_0(t), \ubf_0(t), t)}{\partial \ubf_0(t)}$, and $\Cbf(t) = \frac{\partial h(\xbf_0(t), \ubf_0(t), t)}{\partial \xbf_0(t)}$. Furthermore, $\Abf$, $\Bbf$, and $\Cbf$ are all known matrices that solely depend on the nominal trajectories $\ubf_0(t), \xbf_0(t), \ybf_0(t)$. It is then clear that if the nonlinear system is time-invariant, and if $\ubf_0(t) \equiv \ubf_0$ is constant, and if $\xbf_0(t) \equiv \xbf_0$ is a fixed point, then Eq.~\eqref{eq:ltv-linearized} will take the LTI form (Eq.~\eqref{eq:LTI}). In either case, it is important to remember that this linearization is a valid approximation only locally (in the vicinity of the nominal system), and the original nonlinear system must be studied whenever the system leaves this vicinity.

Leaving the simplicity of linear systems significantly complicates the controllability, observability, and optimal control problems. Fortunately, if the linearization in Eq.~\eqref{eq:ltv-linearized} is controllable (observable), then the nonlinear system is also locally controllable (observable) \cite{EDS:13} (see the topic of linearization validity discussed above). Notably, the converse is not true; the linearization of a controllable (observable) nonlinear system need not by controllable (observable). In such a case, one can take advantage of advanced generalizations of the linear rank condition for nonlinear systems~\cite{EDS:13}, although these tend to be too involved for practical use in large-scale neuronal network models. Interestingly, obtaining optimality conditions for the optimal control of nonlinear systems is not significantly more difficult than that of linear systems. However, solving these optimality conditions (which can be done analytically for linear systems with quadratic cost functions, as mentioned above) leads to non-convex optimization problems that lend themselves to no more than numerical solutions~\cite{DEK:04}.

\section{Models of control \& communication: Areas of distinction, points of convergence}
\label{sec:section4}

In this section, we build upon the descriptions of communication and control provided in Sections \ref{sec:section2} \& \ref{sec:section3} by seeking areas of distinction and points of convergence. We crystallize our discussion around four main topic areas: abstraction versus biophysical realism, linear versus nonlinear models, dependence on network attributes, and the interplay across different spatial or temporal scales. Our consideration of these topics will motivate a discussion of the outstanding challenges and directions for future research, which we provide in Section \ref{sec:section5}. 

\subsection{Abstraction vs. biophysical realism}

Across scientific cultures and domains of inquiry, the requirements of simplicity and tractability place strong constraints on the formulation of theoretical models. Depending on the behavior that the theory aims to capture, the models can capture detailed realistic elements of the system with the inputs from experiments \cite{Bansal2019:Cognitive,Bansal2018:DataDriven}, or the models can be more phenomenological in nature with a pragmatic intent to make predictions and guide experimental designs . An example of a detailed realistic model in the context of neuronal dynamics is the Hodgkin-Huxley model, which takes into account the experimental results from detailed measurements of time-dependent voltage and membrane current\cite{Abbott_1990}. A corresponding example of a more phenomenological model is the Hopfield model, which encodes neuronal states in binary variables. 

\noindent \textbf{Communication models.} Communication models similarly range from the biophysically realistic to the highly phenomenological. Dynamical models informed by empirically measured natural frequencies, empirically measured time delays, and/or empirically measured strengths of structural connections place a premium on biophysical realism \cite{murphy2020multimodal,shirner2018inferring,chaudhuri2015large}. In contrast, Kuramoto oscillator models for communication through coherence can be viewed as less biophysically realistic and more phenomenological \cite{breakspear2010generative}. Communication models also capture the state of a system differently, whether by discrete variables such as on/off states of units, or by continuous variables such as the phases of oscillating units. The diversity present in the current set of communication models allows theoreticians to make contact with experimental neuroscience at many levels \cite{BassettNatrev_2018,Ritter2013:TheVirtua,Kopell2014:Beyond,Sanz-Leon2015:Mathematical}. 

Alongside this diversity, communication models also share several common features. For instance, the state variables chosen to describe the dynamics of the system are motivated by neuronal observations and thus represent the system's biological, chemical, or physical states. The dynamics that state variables follow are also typically motivated by our understanding of the underlying processes, or approximations thereto. In building communication models, the experimental observations and intuition typically precede the mathematical formulation of the model, which in turn serves to generate predictions that help guide future experiments. A particularly good example of this experiment-led theory is the Human Neocortical Neurosolver, whose core is a neocortical circuit model that accounts for biophysical origins of electrical currents generating MEG/EEG signals \cite{jones2020human}. Having been concurrently developed with experimental neuroscience, theoretical models of communication are intricately tied to currently available measurements.

The closeness to biophysical mechanisms is a feature that is also typically shown in other types of communication models. One might think that topological measures devoid of a dynamical model tend to place a premium on phenomenology. But in fact, the cost functions that brain networks optimize typically reflect metabolic costs, routing efficiency, diffusion efficiency, or geometrical constraints \cite{Laughlin_2003,Sporns_2018,Zhou:2020,avena2017path}. Minimization of metabolic costs has been shown to be a major factor determining the organization of brain networks \cite{Laughlin_2003}. Further, such constraints on metabolism also place limits on signal propagation and information processing.

\noindent \textbf{Control models.} Are these features of communication models shared by control models? Control models have their origin in Maxwell's analysis of the centrifugal governor that stabilized the velocity of windmills against disturbances caused by the motions of internal components \cite{Maxwell:1868}. The field of control theory was later further formalized for the stability of motion in linearized systems \cite{routh:1877}. Today, control theory is a framework in engineering used to design systems and to develop strategies to influence the state of a system in a desired manner \cite{Tang_2018}. More recently, the framework of control theory has been applied to neural systems in order to quantify how controllable brain networks are, and to determine the optimal strategies or regions that are best to exert control on other regions \cite{Tang_2018,Gu_optim_trajec:2017,Tang_natcomm:2017}. Although initial efforts have proven quite successful, control theory and more generally, the theory of linear systems, has traditionally concerned itself with finding the mathematical principles behind the design and control of linear systems \cite{KailathLinearSys}, and is applicable to a wide variety of problems in many disciplines of science and engineering. Because the application of control theory to brain networks has been a much more recent effort, identification of appropriate state variables that are best posed to provide insights on control in brain networks is a potential area of future research. 

Applied to brain networks, control theoretical approaches have mostly utilizes detailed knowledge of structural connections while assuming the linear dynamics formulated in Eq.~\ref{eq:LTI-state}. This simplifying abstraction implies that the influence of a system's state at a given time propagates along the paths of the structural network encoded in $\Abf$ of Eq.\ref{eq:LTI-state} to affect the system's state at the next time point. The type of influence studied here is most consistent with the diffusion-based propagation of signals in communication models, and intuitively leads to the expected close relationship between diffusion-based communication measures and control metrics. Indeed such a relationship exists between the impulse response (outlined in the previous section) and the network communicability. We elaborate further on this relationship in the next subsection. 

Some metrics that are commonly used to characterize the control properties of the brain are average controllability, modal controllability, and boundary controllability. These statistical quantities can be calculated directly from the spectra of the controllability Gramian $\Wbf_T$ and the adjacency matrix $\Abf$~\cite{FabioMetrics_2014}. A related and important quantity of interest here is the minimum control energy defined as Eq.~\ref{eq:min-energy} with $\ubf(t)$ denoting the control signals. While this quadratic dependence of `energy' on input signals is appropriate for a linearized description of the system around a steady state, its actual dependence on the exogenous control signals must depend on several details such as the cost of generating control signals and the cost of coupling them to the system. In this sense, the control energy is a relatively abstract concept whose interpretation has yet to be linked to the physical costs of control in brain networks. This observation loops back to the fact that the development of control theory models has been more as an abstract mathematical framework which is then borrowed by several fields and thereafter modified by context. We discuss possible ways of reconciling the cost of control with actual biophysical costs known from communication models in Section \ref{sec:section5}.

\subsection{Linear vs. non-linear models}

In models of communication and dynamics, a reoccurring motif is the propagation of signal along connections. Graph measures such as small-worldness, global efficiency, and communicability assume that strong and direct connections between two neural units facilitate communication \cite{Watts1998,Sporns_2018,Estrada2008}. While these measures capture an intuitive concept and have been useful in predicting behavior, they themselves do not explicitly quantify the mechanism of communication or the form of the information. Dynamical models overcome the former limitation by quantitatively defining the neural states of a system, and encoding the mechanism of communication in the differential or difference equations \cite{Breakspear_2017,Estrada_2012}. However, they only partially address the latter limitation, as it is unclear how a system might change its dynamics to communicate different information. 

There is, of course, no single spatial and temporal scale at which neural systems encode information. At the level of single neurons, neural spikes encode visual \cite{Hubel1959} and spatial \cite{Moser2008} features. At the level of neuronal populations in electroencephalography, changes in oscillation power and synchrony reflect cognitive and memory performance \cite{Klimesch1999}. At the level of the whole brain, abnormal spatio-temporal patterns in functional magnetic resonance imaging reflect neurological dysfunction \cite{Broyd2009}. To accommodate this wide range of spatial and temporal scales of representation, we put forth control models as a rigorous yet flexible framework to study how a neural system might modify its dynamics to communicate.

\subsubsection{Linear models: level of pairwise nodal interactions}

The most immediate relationship between dynamical models and information is through the system's states. From this perspective, the activity or state of a single neural unit is the information to send, and communication occurs diffusively when the states of other neural units change as a result. There is an exact mathematical equivalence between communicability using factorial weights in Eq.~\ref{eq:communicability}, and the impulse response of a linear dynamical system in Eq.~\ref{eq:impulse} through the matrix exponential. Specifically, we realize that the matrix exponential in the impulse response, $e^{\Abf t}$, can be written as communicability with factorial weights, such that
\begin{align*}
e^{\Abf t} = \sum_{k=0}^{\infty} c_k \Abf^k, \hspace{0.5cm} \mathrm{where} \hspace{0.5cm} c_k = \frac{t^k}{k!}.
\end{align*}
This realization provides an explicit link between connectivity, dynamics, and communication \cite{Estrada_2012}. From the perspective of connectivity, the element in the $i$-th row and $j$-th column of the matrix exponential, $[e^{\Abf}]_{ij}$ is the total strength of connections from node $j$ to node $i$ through paths of all lengths. From a dynamic perspective, $[e^{\Abf t}]_{ij}$ is the change in the activity of node $i$ after $t$ time units as a direct result of node $j$ having unit activity. Hence, the matrix exponential explicitly links a structural path-based feature to causal changes in activity under linear dynamics.

\subsubsection{Linear models: level of network-wide interactions}
Increasingly, the field is realizing that the activity of neural systems is inherently distributed at both the neuronal \cite{Steinmetz2019,Yaffe18727} and areal \cite{Tavor2016} levels. Hence, information is not represented as the activity of a single neural unit, but the pattern of activity, or \emph{state}, of many neural units. As a result, we must broaden our concept of communication as the transfer of the system of neural units from an initial state $\xbf(0)$ to a final state $\xbf(t)$. This perspective introduces a rich interplay between the underlying structural features of inter-unit interactions, and the dynamics supported by the structure to achieve a desired state transition.

A crucial question in this distributed perspective of communication is the following: Given that a subset of neural units are responsible for communication, what are the possible states that can be reached? For example, it seems extremely difficult for a single neuron in the human brain to transition the whole brain to any desired state. This exact question has a clear and exact answer in the theory of dynamical systems and control through the \emph{controllability matrix}. Specifically, given a subset of neural units $K$ called the \emph{control set} that are responsible for communication (either of their current state or the external stimuli applied to them) to the rest of the network, the space of possible state transitions is given by weighted sums of the columns of the controllability matrix, i.e., the \emph{controllable subspace} (cf. Section~\ref{subsec:con-obs}). Many studies in control theory are therefore directly relevant for communication, such as determining whether or not a particular control set can transition the system to any state given the underlying connectivity \cite{Lin1974}, or whether reducing the controllable subspace by removing neurons reduces the range of motion \emph{in-vivo} \cite{Yan2017a}.

\subsubsection{Linear models: accounting for biophysical costs}
While determining the theoretical ability of performing a state transition is important, the neural units responsible for control may have to exert a biophysically infeasible amount of effort to perform the transition. Such a constraint is known to be present in many forms such as metabolic cost \cite{liang2013coupling,Laughlin_2003}, and firing rate capacity \cite{sengupta2013balanced}. These constraints are explicitly taken into account in control theory through \emph{minimum energy control}, and by extension, \emph{optimal control}. As detailed in Section~\ref{subsec:con-prac}, the minimum energy control places a homogeneous quadratic cost (control energy) on the amount of effort that the controlling neural units must exert to perform a state transition (Eq.~\ref{eq:min-energy}) while the general LQR optimal control additionally includes the level of activity of the neural units as a cost to penalize infeasibly large states (Eq.~\ref{eq:lqr}). 

Within this framework of capturing distributed communication and biophysical constraints, there remains the outstanding question of how structural connectivity contributes to communication. What features of connectivity enable a set of neural units to better transition the system than another set of units? To this end, many summary statistics have been put forth, mostly in terms of the controllability Gramian (Eq.~\eqref{eq:Cont-LTI-Gram}) due to its crucial role in determining the cost of control (Eq.~\eqref{eq:E_as_gramian}). Among them are the trace of the inverse of the Gramian, $\mathrm{Tr}(\Wbf_T^{-1})$ which quantifies the average energy needed to reach all states on the unit hypersphere (Fig. \ref{fig:Fig3}), and the square root of the determinant of the Gramian $\sqrt{\det(\Wbf_T)}$ (or its logarithm) which is proportional to the volume of states that can be reached with unit input \cite{Muller1972}. Other studies summarize the contribution of connectivity from individual nodes \cite{Simon1968,Gu:2015} or multiple nodes \cite{Pasqualetti2014,JasonKim_2018}, leading to potential candidates for new measures of communication.

\subsubsection{Non-linear models: oscillators and phases}
When faced with the task of studying complex communication dynamics in neural systems, it is evident that the richness of neural behavior extends beyond linear dynamics. Indeed, a typical analysis of neural data involves studying the power of the signals at various frequency bands for behaviors ranging from memory \cite{Klimesch1999} to spatial representations \cite{Moser2008}, underscoring the importance of nonlinear oscillations. To capture these oscillations, the earliest models of Hodgkin and Huxley \cite{Hodgkin_1952}, with subsequent simplifications of Izhikevich \cite{Izhikevich2003} and FitzHugh-Nagumo \cite{Fitzhugh_1961} neurons, as well as population-averaged \cite{Wilson:1972} systems, contain nonlinear interactions that can generate oscillatory behavior. In such systems, how do we quantify information and communication? Further, how would such a system change the flow of communication?   

Some prior work has focused on lead-lag relationships between the signal phases \cite{GN-AZ-VVN-AS-NK-TB-KM:08,CJS-ECWVS:12,Palmigiano_2017}, where the relation implies that communication occurs by the leading unit transmitting information to the lagging unit. A fundamental and ubiquitous equation to model this type of system is the Kuramoto equation (Eq.~\ref{eq:kuramoto}), where each neural unit has a phase $\theta_i(t)$ that evolves forward in time according to the natural frequency $\omega_i$ and a sinusoidal coupling with the phases of the other units $\theta_j$, weighted by the coupling strength $A_{ij}$ \cite{YK:03,Acebron:2005}. This model has a vast theoretical and numerical foundation with myriad applications in control systems \cite{Dorfler2014}.

Given an oscillator system with fixed parameters, how can the system establish and alter its lead-lag relationships? In the regime of \emph{frequency synchronization} where the natural frequencies are not identical, the oscillators converge to a common synchronization frequency $\omega_{\mathrm{sync}}$. As a result, the relative phases with respect to this frequency remain fixed at $\theta_{\mathrm{sync}}$ \cite{Dorfler2014}, thereby establishing a lead-lag relationship. In this regime, the nonlinear oscillator dynamics can be linearized about $\omega_{\mathrm{sync}}$, to generate a new set of dynamics
\begin{align*}
\dot{\theta}_i \approx \omega_i - \sum_{j=1}^N L_{ij}\theta_j,
\end{align*}
where $\Lbf$ is the network Laplacian matrix of the coupling matrix $\Abf$. In \cite{Skardal2015}, the authors begin with an unstable general oscillator network that is not synchronized (i.e., does not have a true $\theta_{\mathrm{sync}}$, and perform state-feedback to stabilize an unstable set of phases $\thetabf^*$, thereby inducing frequency synchronization with the corresponding lead-lag relationships. The core concept behind this state-feedback is to designate a subset of oscillators as ``driven nodes,'' and add an additional term that modulates the phases of these oscillators according to
\begin{align*}
    \dot{\theta}_i = \underbrace{\omega_i + \sum_{j=1}^N A_{ij}\sin(\theta_j-\theta_i)}_{\mathrm{natural~dynamics}} + \underbrace{F_i\sin(\theta_i^*-\theta_i)}_{\mathrm{state-feedback}}.
\end{align*}
Subsequent work focuses on expanding the form of the control input \cite{Skardal2016}, and modifications to the coupling strength to a single node \cite{Fan2019}. Hence, we observe that targeted modification to the dynamics of subsets of oscillators can indeed set their lead-lag relationships. 

Generally, oscillator systems are not inherently phase oscillators. For example, the Wilson-Cowan \cite{Wilson:1972}, Izhikevich \cite{Izhikevich2003}, and FitzHugh-Nagumo \cite{Fitzhugh_1961} models are all oscillators with two state variables coupled through a set of nonlinear differential equations. The transformation of these state variables and equations into a phase oscillator form is the subject of weakly-coupled oscillator theory \cite{Dorfler2014,HGS-PW:90}. In the event that the oscillators are not weakly coupled, then controlling the dynamics and phase relations begins to fall under the purview of linear time-varying systems (Eq.~\ref{eq:ltv}) and nonlinear control \cite{HKK:02,EDS:13}.

\subsection{Dependence on network attributes}
\label{sec:int-seg}
In network neuroscience, recent studies have begun to characterize how network attributes influence communication and control in neuronal and regional circuits. In neuronal circuits, the spatiotemporal scale of communication has been studied from the perspective of statistical mechanics in the context of neuronal avalanches \cite{Beggs_2003}. Such studies show that activity propagates in a critical \cite{Beggs:2012}, or at least slightly subcritical \cite{Priesemann:2014, Wilting:2018}, regime. In a critical regime, the network connections are tuned to optimally propagate information throughout the network \cite{Beggs_2003}. Studies of microcircuits also show more explicitly that certain network topologies can play precise roles in communication. Hubs, which are neural units with many connections, often serve to transmit information within the network \cite{Timme:2016}. Groups of such hubs are called rich-clubs \cite{towlson2013rich,colizza2006detecting}, which have been observed in a wide range of organisms \cite{Shimono:2014, Faber:2019}, and they dominate information transmission and processing in networks \cite{Faber:2019}.

Cortical network topologies have highly non-random features \cite{Song:2005}, which may support more complex routing of communication \cite{avena2019spectrum}. In studies of neuronal gating, one group of neurons, such as the mediodorsal thalamus, can either facilitate or inhibit pathways of communication, such as that from the hippocampus to the prefrontal cortex \cite{Floresco:2003}. Such complex routing of communication requires non-linear dynamics, such as shunting inhibition \cite{Borg-Graham:1998}. Some models simulate inhibitory dynamics on cortical network topologies to study how those topologies may support the complex communication dynamics that occur in visual processing, such as visual attention \cite{Anderson:1987,Olshausen:1993}.

Points of convergence between communication and control have been observed in regional brain networks. For example, hubs are studied in functional covariance networks and structural networks. Structural hubs are thought to act as sinks of early messaging accelerated by shortest-path structures and sources of transmission to the rest of the brain \cite{Misic_2015}. The highly connected hub's connections may support both the average controllability of the brain as well as the brain's robustness to lesions of a fraction of the connections \cite{Lee_2019}. An area of distinction between control and communication in brain networks may depend on the hub topology. While communication may depend on the average controllability of hubs to steer the brain to new activity patterns, the brain regions that steer network dynamics to difficult-to-reach states tend to not be hubs \cite{Gu:2015}. In determining the full set of nodes that can confer full control of the network, hubs tend to not be in this set of driver nodes \cite{Liu2011a}. A point of convergence between communication and control is the consideration of how the brain network broadcasts control signals. Whereas the high-degree of hubs may efficiently broadcast integrated control signals across the brain network in order to steer the brain to new patterns of activity, the brain regions with lower degree may receive a greater rate of control signals which are then transmitted to steer the brain to difficult-to-reach patterns of activity \cite{Zhou:2020}.

To strike a balance between efficiency, robustness, and diverse dynamics, brain networks may have evolved towards optimizing brain network structures supporting and constraining the propagation of information. Brain networks reach a compromise between routing and diffusion of information compared to random networks optimized for either routing or diffusion \cite{Sporns_2018}. Brain networks also appear optimized for controllability and diverse dynamics compared to random networks \cite{Tang_natcomm:2017}. To understand how the brain can circumvent trade-offs between objectives like efficiency, robustness, and diverse dynamics, future studies could further investigate the network properties of the spectrum of random networks optimized toward these objectives. Existing studies focus on the trade-off between two objectives, such as network structure supporting information routing or diffusion, or average versus modal controllability. However, multi-objective optimization allows for further investigation of Pareto-optimal brain network evolution towards an arbitrarily large set of objective functions \cite{avena2015network,avena2014using}.

The convergence between communication and control exists largely via the network topologies with which they are related. Given the importance of `rich-club hubs' and similar topological attributes in integration and processing of information, it is natural to ask if similar properties also contribute to controllability or observability properties in brain networks. More specifically, given a region in the brain network with specific topological properties such as high degree, betweenness centrality, closeness centrality, or location between different modules, what is the relationship between its role in information integration or processing and its role in controllability and observability? The tri-faceted interface of communication, control, and network topology holds great possibilities for future work, and some recent efforts have begun to relate the three \cite{Ju:2018}.
\\

\subsection{Interplay of multiple spatiotemporal scales}

Most complex systems exhibit phenomena at one spatiotemporal scale that depend upon phenomena occurring at another spatiotemporal scale. This interplay of scales is evident, for example, in the hierarchical energy cascade from lower modes (larger length scales) to higher modes (smaller length scales) in turbulent fluids \cite{Frisch_1995}, multiscaled models of morphogenesis \cite{Jensen_2015, Newman_2008, Walpole_2013}, and multiscaled models of cancer \cite{Deisboeck_2011}. A convenient way to study such an interplay is to transform the variables of mathematical models to their corresponding Fourier conjugate variables. This approach serves to map the larger length scales to Fourier modes of smaller wave-lengths, and to map the longer time-scales to smaller frequency bands. In most complex systems, current research efforts seek a quantitative understanding of the interwoven nature of different spatiotemporal scales, which in turn can lead to an understanding of the system's emergent behavior.

\noindent \textbf{Communication models.} As one of the most complex systems, the brain naturally exhibits a rich cross-talk between different spatiotemporal scales. A key example of interplay among \emph{spatial} scales is provided by recent evidence that activity propagates in a slightly subcritical regime, in which activity ``reverberates'' within smaller groups of neurons while still maintaining communication across those groups \cite{Wilting:2018}. That cross-talk is structurally facilitated by topological features characteristic of each spatial scale: from neurons to neuronal ensembles to regions to circuits and systems \cite{Shimono:2014,Bansal2019:Cognitive,Kopell2014:Beyond}. A key example of interplay among \emph{temporal} scales is cross frequency coupling \cite{Canolty_2010}, which first builds upon the observation that the brain exhibits oscillations in different frequency bands thought to support information integration in cognitive processes from attention to learning and memory \cite{Cannon2014:Neurosystems,Basar_2004, Basar_2001, breakspear2010generative,Kopell2010:GammaAndTheta,Kopell2010:AreDifferentRhythms}. Cross frequency coupling can occur between region $i$ in one frequency band and region $j$ in another frequency band, and be measure statistically \cite{Tort_2010}. The phenomenon is thought to play a role in integrating information across multiple spatiotemporal scales \cite{Aru_2015}. For example, directional coupling between hippocampal $\gamma$ oscillations and the neocortical $\alpha$/$\beta$ oscillations occurs in the context of episodic memory \cite{Griffiths_2019}. Interestingly, anomalies of oscillatory activity and cross frequency coupling can serve as biomarkers of neuropsychiatric disease \cite{Basar_2013}.

While cross-scale interactions exist, most biophysical models have been developed to address the dynamics evident at a single scale. Despite that specific goal, such models can also sometimes be used to better understand the relations between scales. For example, the notable complexity present at small scales often gives way to simplifying assumptions in some limits. That mathematical characteristic allows for coarse grained models to be derived at the next larger spatiotemporal scale \cite{Deco_2008,Breakspear_2017}. Key examples of such coarse-graining include (i) the derivation of the Fokker-Planck equations for neuronal ensembles in the limit where the firing activity of individual neurons are independent processes, and (ii) the derivation of neural mass models in the limit of strong coherence in neuronal ensembles. The procedure of coarse-graining is thus one theoretical tool that helps to bridge mathematical models of the system at different spatial scales, in at least some simplifying limits.

Communication models also allow for an interplay between different length scales by two other mechanisms: (i) the inclusion of non-linearities, which allow for coupling between different modes, and (ii) the presence of global constraints. Regarding the former mechanism, a linearized description of dynamics can typically be transformed into the `normal' modes of the system (i.e., the eigenvalues of the $\Abf$ matrix in Eq.~\eqref{eq:LTI}), and this description does not allow for the inter-mode coupling that would otherwise be permissible in a nonlinear communication model. As an example of such nonlinear models, neural mass models and Wilson-Cowan oscillators can exhibit cross frequency coupling via the mechanism of phase-amplitude coupling \cite{Onslow_2014,EN-JC:19-acc,Daffertshofer2011:OnTheInfluence}, where the amplitude of high-frequency oscillations are dependent on the phase of slowly-varying oscillations. Regarding the latter mechanism, inter-regional communication is constrained by the global design of brain networks which have evolved under constraints on transmission speed, spatial packing, metabolic cost, and communication efficiency \cite{Zhou:2020,Laughlin_2003}. 

\noindent \textbf{Control models.}
Are control models -- like communication models -- conducive to a careful study of the rich interplay of multiple spatiotemporal scales in neural systems? This question may be particularly relevant when control signals can only be injected at a given scale while the desired changes in brain activity lie at an altogether different scale. One way to interlink local and global scales in control models is to use global constraints, just as we discussed in the context of communication models. The application of control theory to a given system often requires finding control signals that minimize the overall cost of control and/or that constrain the system to follow an optimum trajectory. Both of these goals can be recast in terms of optimization problems where a suitable cost function is minimized (see Section \ref{sec:section3}). In this sense, the global constraints dictate the control properties of the system. 

Given that linear models produce a limited number of behaviors in solution-space and do not allow for coupling between different modes (as discussed above and in Section \ref{sec:section3}), the application of nonlinear control theory is highly warranted to bring an interesting interplay between different scales. Here, the theory of singular perturbations~\cite{HKK:02} provides a natural and powerful tool to formally characterize the relationship between temporal scales in a multiple-timescale system~\cite{EN-JC:18-tacII}. This theory formalizes the intuition that with respect to a subsystem at a `medium' (or reference) temporal scale, the activity of subsystems at slower temporal scales is approximately constant while the activity of those at faster timescales can be approximated by their attractors (hence neglecting fast transients), and is particularly relevant for brain networks whereby timescale hierarchies have been robustly observed, both \emph{in vivo} \cite{JDM-AB-DJF-RR-JDW-XC-CP-TP-HS-DL-XW:14} and using computational modeling~\cite{RC-KK-MG-HK-XW:15}. Such extended control models thus form a natural approach toward a careful evaluation of cross-scale interactions.

Another concrete way in which multiple scales can be incorporated into control models -- while retaining simple linearized dynamics -- is to build the complexity of the system into the network representation of the brain itself. The formalism of multilayered networks allows for the complexity of interacting spatiotemporal scales to be built into the structure where the layer identification (and the definition of interlayer and intralayer edges) can be based upon the inherent scales of the system. One concrete example of this architecture is a two-layer network in which each layer shares the same nodes (brain regions) but represents different types of connections. One such two-layer network could have nodes representing brain regions, edges in one layer representing slowly-varying structural connections, and edges in the second layer representing functional connectivity with faster dynamics. Moreover, different frequency bands can be explicitly identified as separate layers in a multiplex representation of brain networks \cite{Buldu_2017}, allowing for a careful investigation of cross-frequency coupling. It would be of great interest in future to combine such a multilayer representation with the simple LTI dynamics in control models to better understand how control signals can drive desired (multiscale) dynamics. The inbuilt complexity of structure can thus partially compensate for the requirement of dynamical complexity, and thereby increase the validity of LTI dynamics when applied to brain networks.

\section{Outstanding challenges \& future directions}
\label{sec:section5}

\begin{figure*}
\includegraphics[scale=0.24]{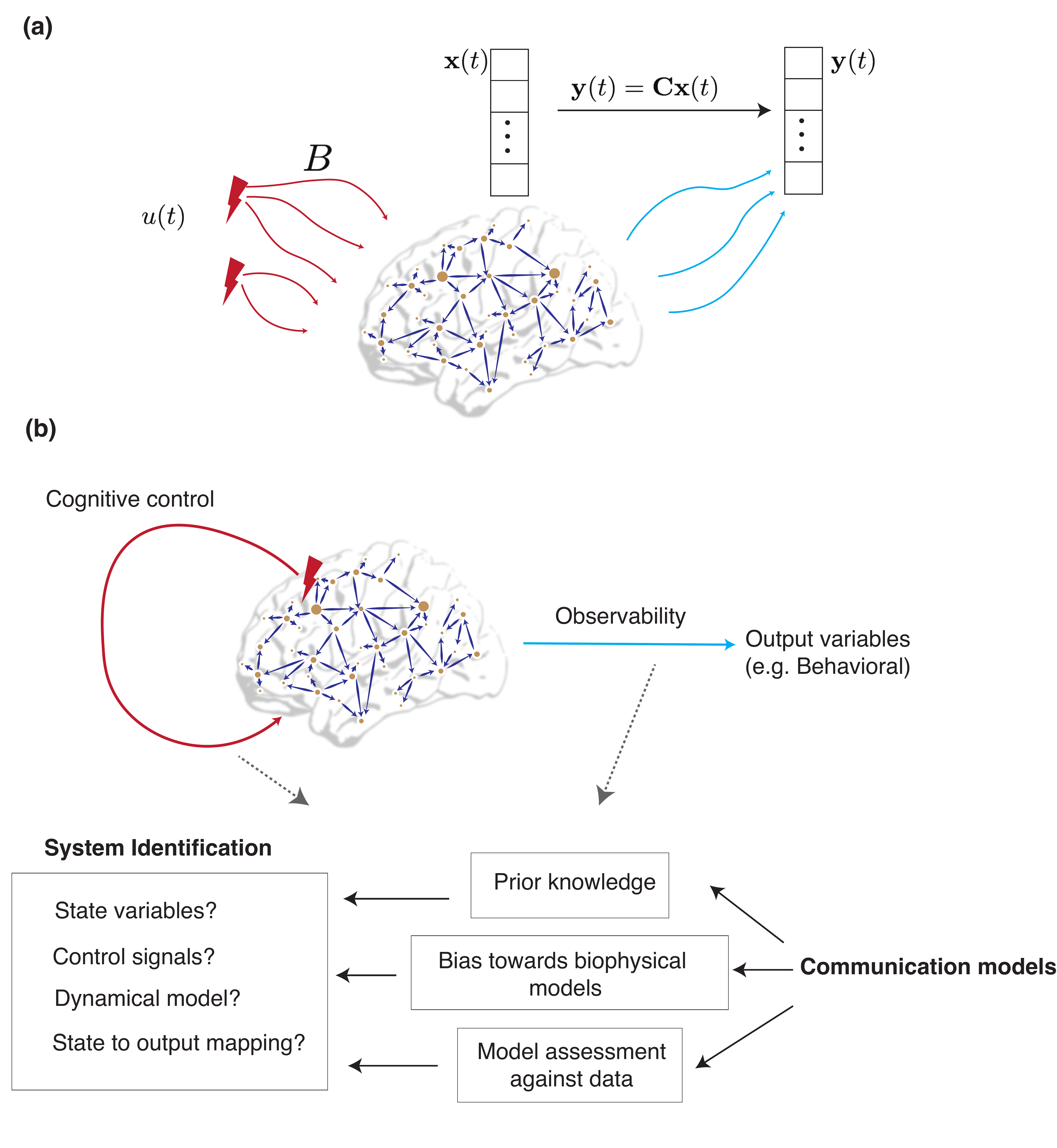}
\begin{flushleft}
\caption{\textbf{Intersection of control theory, communication models and system identification.} (a) Exogenous input applied to brain networks leads to the evolution of brain states (represented by vector $\xbf(t)$) according to a specified dynamics (see Section~\ref{sec:section3}). Given the observed vector $\ybf(t)$, the observability of brain states is determined by the invertibility of the state-to-output map (denoted by the matrix $\mathbf{C}$ ). (b) \emph{System identification} has been proposed as a key step in the application of network control theory to understand cognitive control and other cognitive processes. The key steps of the system identification process can be further integrated with the insights from communication models, thus guiding future research on the formulation of theoretical frameworks to understand cognition.
}
\label{fig:Fig5}
\end{flushleft}
\end{figure*}

Having discussed several areas of distinction and points of convergence, in this section we turn to a discussion of outstanding challenges and directions for future research. We focus our discussion around three primary topic areas: observability and information representation, system identification and causal inference, and biophysical constraints. We will close in Section~\ref{sec:section6} with a brief conclusion. 

\noindent
\subsection{Observability and information representation}
In the theory of linear systems, observability is a notion that is dual to controllability and is considered on an equal footing~\cite{KailathLinearSys} (cf. Section~\ref{subsec:con-obs}). Interestingly, however, this equality has not been reflected in the use of these notions to study neural systems. The controllability properties of brain networks have comprised a large focus of the field, whereas the concept of observability has not been applied to brain networks to an equivalent extent. The focus on the former over the latter is likely due in large part to historical influences over the processes of science, and not to any lack of utility of observability as an investigative notion. Indeed, observability may be crucial in experiments where the state variables differ from the variables that are being measured. A canonical example is situations where the state variables correspond to the average firing rates of different neuronal populations whereas the outputs being measured are behavioral responses. More precisely, specific stimuli (control signals) can be represented to have a more direct effect on neuronal activity patterns (state variables) which, in turn, produce behavioral responses such as eye movements (output variables) after undergoing cognitive processes in brain. In this example, observability refers to the ability of a system model to allow its underlying state (neuronal activity) to be uniquely determined based on the observation of samples of its output variables (behavioral responses) and an appropriate estimation method. Along similar direction, optimal control based methods have been applied to detect the clinical and behavioral states and their transitions \cite{Sabato2011:optimalstatedetection,Sabato2012:statedetection}. 

As discussed at length in Section~\ref{sec:section3}, the observability of state variables depends on the mapping between state variables and output variables encoded and determined by a state-to-output mapping (i.e., the matrix $\Cbf$ in Eq.~\eqref{eq:LTI}). In this regard, the determination of state variables from measured output variables is a problem that, in spirit, bears resemblance to the well-studied problems of neural encoding and decoding of information. While the process of neural encoding involves representing the information about stimuli in the spiking patterns of neurons, the process of neural decoding is the inverse problem of determining the information contained in those spiking patterns to infer the stimuli~\cite{churchland2012neural}. Detailed statistical methods and computational approaches have been developed to address these problems~\cite{kao2014information}. The field of neuronal encoding and decoding stands at the interface of statistics, neuroscience, and computer science, but has not previously been strongly linked to control theoretic models. Nevertheless, such a link seems intuitively fruitful, as the problem of determining state variables from a measured output and the problem of determining stimuli from the measured spiking activity of neurons are conceptually quite similar to one another~\cite{chen2015advanced}.

In the field of control theory, analogous problems are generally referred to under the umbrella of \emph{state estimation} and \emph{filtering}. For example, the Kalman filter in its simplest form consists of a recursive procedure to compute an optimal estimate of the state given the observations of inputs and outputs of a linear dynamical system affected by normally distributed noise~\cite{KailathLinearSys}. The conceptual similarity between neuronal decoding and the notion of observability promises to open an interface between control models and the field of neuronal coding. For example, it will be interesting to ask if the tools and approaches from the well-established field of neuronal decoding can be adapted to the framework of control theory and inform us about the observability of internal states of the brain. Framing and addressing such questions will be instrumental in providing insights to the nature of brain states and the dynamics of transitions between them. This intersection is also a potential area to integrate control and communication models, with the goal of generating observed spiking patterns given a set of stimuli. Such an effort could provide a mechanistic understanding of the nature of information propagated during various cognitive tasks, and of precisely how signals are transformed in that process.\\

\noindent
\subsection{System identification and causal inference}
Network control theory promises to be an exciting ground to study and understand intrinsic human capacities such as cognitive control \cite{Gu:2015,cornblath2019sex,Tang_2018,cui2020optimization,Medaglia_2018}. Cognitive control refers to the ability of the brain to influence its behavior in order to perform specific tasks. Common manifestations of cognitive control include monitoring the brain's current state, calculating the difference from the expected behavior for the specific task at hand, and deploying and dynamically adjusting itself according to the system's performance \cite{miller2001integrative}. While cognitive control shares some common features with the theory of network control, the outstanding problem in formalizing that relationship with greater biological plausibility falls primarily within the realm of \emph{system identification}~\cite{Ljung:1987} (Fig. \ref{fig:Fig5}).

System identification is a formal procedure which involves determining appropriate models, variables, and parameters to describe system observations. The key ingredients of a system identification scheme are (a) the input-output data, (b) a family of models, (c) an algorithm to estimate model parameters, and (d) a method to assess models against the data~\cite{Ljung:1987}. A successful system identification scheme applied to a human capacity like cognitive control can lead to a better identification of state variables and controllers and help to bridge the gap between cognitive processes and network control theory. It is here, at the intersection of cognitive control and network control theory, that communication models can again prove to be relevant. Since communication models have investigated state variables and dynamics that are typically relatively close to the actual biophysical description of the system, system identification can benefit from communication models in supplying \emph{prior knowledge}, assigning weights to plausible models, and setting the assessment criterion.

Closely associated with the problem of system identification is the topic of causal inference, which seeks to produce models that can predict the effects of external interventions on a system~\cite{pearl2009causality}. Such an association stems from the fact that dynamical models are intended to quantify how the system reacts to the application of external control inputs (i.e., interventions). In particular, as discussed in Section~\ref{sec:section3}, a controllable model implies the existence of a sequence of external inputs that is able to drive the system to any desired state. Therefore, appropriate control models are expected to express valid causal relationships between the external inputs and their influence on the system state.

System identification methods have been traditionally based on statistical inference methodologies that are concerned with capturing statistical associations (i.e., correlations and dependencies) over time which do not necessarily imply cause-effect relationships~\cite{koller2009probabilistic}. Within that perspective, system identification methods have been most successful in disciplinary areas where the fundamental mechanistic principles across variables (and hence their causal structure) are known, to a large extent, \emph{a priori} (e.g., \emph{white} and \emph{gray} models. Consequently, when considering complex systems such as the brain, which are often associated with high-dimensional measurements potentially affected by hidden variables, the limitations of such methods become relevant, and the models thus produced may need to be further evaluated for their causal validity. In this respect, the intersection of causal inference and (complex) system identification is likely to become a promising area of future research. For example, it will be interesting to see how tools from system identification may evolve to incorporate new methodologies from the theory of causal inference, and how the resulting tools might generate additional requirements for experimental design and data collection in neuroscientific research.

\noindent
\subsection{Biophysical constraints}
In network control models, it is unknown how mathematical control energy relates to measurements of biophysical costs (also see Section~\ref{sec:section4}). Although predictions of control energy input have been experimentally supported by brain stimulation paradigms \cite{stiso2019white, khambhati2019functional}, the control energy costs of the endogenous dynamics of brain activity are not straightforwardly described by external inputs. According to brain network communication models of metabolically efficient coding \cite{Zhou:2020}, an intuitive hypothesis is that the average size of the control signals required to drive a brain network from an initial state to a target state correlates with the regional rates of metabolic expenditure.

Similar questions aiming to discover biophysical mechanisms of cognitive control have been tackled by ongoing investigations of cognitive effort, limited cognitive resources, motivation, and value-guided decision-making \cite{kool2018mental, shenhav2017toward}. However, there is limited evidence of metabolic cost operationalized as glucose consumption as a main contributor to cognitive control. Rather, the dynamics of the dopamine neurotransmitter, transporters, and receptors appear to be crucial \cite{cools2016costs, westbrook2016dopamine}. Recent work in network control theory has provided converging evidence for a relationship between dopamine and control in cognition and psychopathology \cite{braun2019brain}. The subcortical dopaminergic network and frontoparietal cortical network may support the computation and communication of reward prediction errors in models of cost-benefit decision-making, expected value of control, resource rational analysis, and bounded optimality \cite{westbrook2016dopamine}. 

Cognitive control theories distinguish between the costs and allocation of control \cite{shenhav2017toward}. Costs include behavioral costs, opportunity costs, and intrinsic implementation costs. Prevailing proposals of how the brain system allocates control include depletion of a resource, demand on a limited capacity, and interference by parallel goals and processes. Control allocation is then defined as the expected value of control combined with the intrinsic costs of cognitive control. Broadly, a control process consists of monitoring control inputs and changes, specifying how and where to allocate control, and regulating the transmission of control signals~\cite{AO:14,THS-JL:14,EN-FP-JC:19-jcn}. Notably, the implementation of how the brain regulates the transmission of control signals and accounts for the intrinsic costs of cognitive control require further development, providing promising avenues to apply mathematical models of brain network communication and control. Existing control models of brain dynamics, for instance, have mostly assumed noise-free dynamics (but also see~\cite{schiff2012neural,ZH-SVS:17}). Recent communication models can be applied to model noisy control by defining how brain regions balance communication fidelity and signal distortion in order to efficiently transmit control input messages at an optimal rate to receiver brain regions with a given fidelity \cite{Zhou:2020}. Such an approach may be particularly fruitful in ongoing efforts seeking to better understand the relations between cognitive function, network architecture, and brain age both in health and disease \cite{bunge2012brain,morgan2018network}.\\

\section{Conclusion}
\label{sec:section6}

The human brain is a complex dynamical system whose functions include both communication and control. Understanding those functions requires careful experimental paradigms and thoughtful theoretical constructs with associated computational models. In recent years, separate streams of literature have been developed to formalize the study of communication in brain networks, as well as to formalize the study of control in brain networks. Although the two fields have not yet been fully interdigitated, we posit that such an integration is necessary to understand the system which produces both functions. To support future efforts at their intersection, we briefly review canonical types of communication models (dynamical, topological, and information theoretic), as well as the formal mathematical framework of network control (controllability, observability, linear system control, linear time-varying system control, and non-linear system control). We then turn to a discussion of areas of distinction between the two approaches, as well as points of convergence. That comparison motivates new directions in better understanding the representation of information in neural systems, in using such models to make causal inferences, and in experimentally probing the biophysical constraints on communication and control. Our hope is that future studies of this ilk will provide fundamental, theoretically grounded advances in our understanding of the brain.

\section{Citation Diversity Statement}

Recent work in neuroscience and other fields has identified a bias in citation practices such that papers from women and other minorities are under-cited relative to the number of such papers in the field \cite{Dworkin2020.01.03.894378, maliniak2013gender, caplar2017quantitative, chakravartty2018communicationsowhite, YannikThiemKrisF.SealeyAmyE.FerrerAdrielM.Trott2018, dion2018gendered}. Here we sought to proactively consider choosing references that reflect the diversity of the field in thought, race, geography, form of contribution, gender, and other factors. We used automatic classification of gender based on the first names of the first and last authors \cite{Dworkin2020.01.03.894378}, with possible combinations including male/male, male/female, female/male, and female/female. Excluding self-citations to the first and senior author of our current paper, the references contain $51.9 \%$ male/male, $16.7\%$ male/female, $19.6\%$ female/male, $8.8 \%$ female/female, and $3\%$ unknown categorization (codes in \cite{github_dale} were used to estimate these numbers). We look forward to future work that could help us to better understand how to support equitable practices in science.

\section{Acknowledgements}
This work was primarily supported by the National Science Foundation (BCS-1631550), the Army Research Office (W911NF-18-1-0244), and the Paul G. Allen Family Foundation. We would also like to acknowledge additional support from the John D. and Catherine T. MacArthur Foundation, the Alfred P. Sloan Foundation, the ISI Foundation, the Army Research Laboratory (W911NF-10-2-0022), the Army Research Office (Bassett-W911NF-14-1-0679, Grafton-W911NF-16-1-0474, DCIST-W911NF-17-2-0181), the Office of Naval Research, the National Institute of Mental Health (2-R01-DC-009209-11, R01-MH112847, R01-MH107235, R21-M MH-106799), the National Institute of Child Health and Human Development (1R01HD086888-01), National Institute of Neurological Disorders and Stroke (R01 NS099348), and the National Science Foundation (BCS-1441502, BCS-1430087, and NSF PHY-1554488). The content is solely the responsibility of the authors and does not necessarily represent the official views of any of the funding agencies.

\bibliography{Comm_review_updated}
\bibliographystyle{ieeetr}

\end{document}